\documentclass{article}
\usepackage{epsfig}

\date{\today}

\usepackage{amsmath}
\usepackage{amssymb}
\usepackage[hang,nooneline,scriptsize]{subfigure}
\begin{document}

\title{Formation of scalar hair on Gauss-Bonnet solitons and black holes}

\author{{\large Yves Brihaye \footnote{email: yves.brihaye@umons.ac.be}
}$^{(1)}$, 
{\large Betti Hartmann \footnote{email:
b.hartmann@jacobs-university.de}}$^{(2)}$
and
{\large Sardor Tojiev \footnote{email:
s.tojiev@jacobs-university.de}}$^{(2)}$
\\ \\
$^{(1)}${\small Physique-Math\'ematique, Universite de
Mons-Hainaut, 7000 Mons, Belgium}\\ 
$^{(2)}${\small School of Engineering and Science, Jacobs University Bremen,
28759 Bremen, Germany}  }

\date{\today}
\setlength{\footnotesep}{0.5\footnotesep}
\newcommand{\dd}{\mbox{d}}
\newcommand{\tr}{\mbox{tr}}
\newcommand{\la}{\lambda}
\newcommand{\ka}{\kappa}
\newcommand{\f}{\phi}
\newcommand{\vf}{\varphi}
\newcommand{\F}{\Phi}
\newcommand{\al}{\alpha}
\newcommand{\ga}{\gamma}
\newcommand{\de}{\delta}
\newcommand{\si}{\sigma}
\newcommand{\bomega}{\mbox{\boldmath $\omega$}}
\newcommand{\bsi}{\mbox{\boldmath $\sigma$}}
\newcommand{\bchi}{\mbox{\boldmath $\chi$}}
\newcommand{\bal}{\mbox{\boldmath $\alpha$}}
\newcommand{\bpsi}{\mbox{\boldmath $\psi$}}
\newcommand{\brho}{\mbox{\boldmath $\varrho$}}
\newcommand{\beps}{\mbox{\boldmath $\varepsilon$}}
\newcommand{\bxi}{\mbox{\boldmath $\xi$}}
\newcommand{\bbeta}{\mbox{\boldmath $\beta$}}
\newcommand{\ee}{\end{equation}}
\newcommand{\eea}{\end{eqnarray}}
\newcommand{\be}{\begin{equation}}
\newcommand{\bea}{\begin{eqnarray}}

\newcommand{\ii}{\mbox{i}}
\newcommand{\e}{\mbox{e}}
\newcommand{\pa}{\partial}
\newcommand{\Om}{\Omega}
\newcommand{\vep}{\varepsilon}
\newcommand{\bfph}{{\bf \phi}}
\newcommand{\lm}{\lambda}
\def\theequation{\arabic{equation}}
\renewcommand{\thefootnote}{\fnsymbol{footnote}}
\newcommand{\re}[1]{(\ref{#1})}
\newcommand{\R}{{\rm I \hspace{-0.52ex} R}}
\newcommand{\N}{{\sf N\hspace*{-1.0ex}\rule{0.15ex}%
{1.3ex}\hspace*{1.0ex}}}
\newcommand{\Q}{{\sf Q\hspace*{-1.1ex}\rule{0.15ex}%
{1.5ex}\hspace*{1.1ex}}}
\newcommand{\C}{{\sf C\hspace*{-0.9ex}\rule{0.15ex}%
{1.3ex}\hspace*{0.9ex}}}
\newcommand{\eins}{1\hspace{-0.56ex}{\rm I}}
\renewcommand{\thefootnote}{\arabic{footnote}}
 \maketitle
\begin{abstract} 
We discuss the formation of scalar hair on Gauss-Bonnet solitons and black holes
in 5-dimensional Anti-de Sitter (AdS) space-time.
We present new results on the static case and point out further details. We find that
the presence of the Gauss-Bonnet term has an influence on the pattern of soliton solutions
for small enough values of the electric charge. 
We also discuss rotating Gauss-Bonnet black holes with and without scalar hair.

\end{abstract}

\section{Introduction}
Most theories of quantum gravity need more than the standard four space-time dimensions to be consistent.
String theory is an example of this. The low energy effective action of these models
in general reproduces Einstein gravity, but also contains terms that are higher order in
the curvature invariants \cite{zwiebach}. In five space-time dimensions this is the 
Gauss-Bonnet
(GB) term, which has the property
that the equations of motion are still second order in derivatives of the metric
functions. Since black holes are thought to be the testing grounds for quantum gravity models it is
of course of interest to study the generalisations of known black hole solutions to include
higher order curvature corrections.  As such explicit
spherically symmetric and asymptotically flat
black hole solutions in GB gravity are known for the uncharged case 
\cite{deser,Wheeler:1985nh}, for the charged case 
\cite{wiltshire} as well as in Anti-de Sitter (AdS)
\cite{Cai:2001dz,Cvetic:2001bk,Cho:2002hq,Neupane:2002bf}
and de Sitter (dS) space-times \cite{Cai:2003gr}, respectively. In
most cases, black holes not only with
spherical ($k=1$), but also with flat ($k=0$) and hyperbolic ($k=-1$) horizon
topology have been considered.
Moreover, the thermodynamics of these black holes has been studied in detail
\cite{Cho:2002hq,Neupane:2002bf,Neupane:2003vz}
and the question of negative entropy for certain GB black holes in dS and AdS
has been discussed \cite{Cvetic:2001bk,Clunan:2004tb}.

One of the most important results of String theory is surely the
AdS/CFT correspondence \cite{ggdual, adscft} that states that a gravity theory in $(d+1)$-dimensional AdS space-time
is equivalent to a conformal field theory (CFT) on the $d$-dimensional boundary of AdS. This
correspondence is a weak-strong coupling duality and can be used to describe
strongly coupled Quantum Field Theories on the boundary of AdS by weakly coupled
gravity theories in the AdS bulk. An application of these ideas is the
description of high temperature superconductivity  with the help of black holes and solitons in
AdS space--time \cite{gubser,hhh,horowitz_roberts,reviews}. In most cases $(3+1)$-dimensional
solutions with planar horizons ($k=0$) were
chosen to account for the fact that high temperature superconductivity is mainly
associated to 2-dimensional layers within the material. 
The basic idea is that at low temperatures a planar black hole in asymptotically
AdS becomes unstable to the condensation
of a charged scalar field since its effective mass
drops below the Breitenlohner-Freedman (BF) bound \cite{bf} for sufficiently low
temperature of the black hole hence
spontaneously breaking the U(1) symmetry.
It was shown that this corresponds to a conductor/superconductor phase
transition. Alternatively, solitons in AdS become unstable to scalar hair formation
if the value of the chemical potential is large enough. Since solitons do not have a 
temperature associated to them this has been interpreted as a zero temperature phase transition between 
an insulator
and a superconductor. 
Interestingly, there seems to be
a contradiction between the holographic superconductor approach and the
Coleman-Mermin-Wagner theorem \cite{CMW} 
which forbids spontaneous symmetry
breaking in $(2+1)$ dimensions at finite temperature. Consequently, it has been
suggested that 
higher curvature corrections and in particular GB terms should 
be included on the gravity side and holographic GB superconductors in $(3+1)$
dimensions have been
studied \cite{Gregory:2009fj}. However, though the critical temperature gets
lowered
when including GB terms, condensation cannot be suppressed -- not even when
backreaction of the space-time is included \cite{Brihaye:2010mr,Barclay:2010up,Siani:2010uw}.

Next to the stability of solutions with flat sections and their application in the context of holographic 
superconductors it is of course also of interest
to discuss the stability of black holes with spherical ($k=1$) or hyperbolic ($k=-1$) 
horizon topology in AdS space-time.

In \cite{Dias:2010ma} the question of the condensation of a uncharged scalar
field on uncharged black holes in $(4+1)$ dimensions has been addressed.
As a toy model for the rotating case, static black holes with hyperbolic
horizons ($k=-1$) were discussed.
In contrast to the uncharged, static black holes with flat ($k=0$) or spherical
($k=1$) horizon topology
hyperbolic black holes possess an extremal limit with near-horizon geometry
AdS$_2\times H^3$ \cite{Robinson:1959ev,Bertotti:1959pf,Bardeen:1999px}. 
This leads to the instability of these black holes with respect to scalar hair formation in the near-horizon
geometry 
as soon as the scalar field mass becomes smaller than the $2$-dimensional
BF bound. Note that these black holes are still asymptotically AdS as long as the $(4+1)$-dimensional
BF bound is fulfilled. These studies were extended to include Gauss-Bonnet corrections \cite{hartmann_brihaye3}.

In \cite{dias2}
static, spherically symmetric $(k=1)$ black hole and soliton solutions to Einstein-Maxwell 
theory coupled to a charged, massless scalar field 
in $(4+1)$-dimensional global AdS space-time have been studied. The existence
of solitons in global AdS was discovered in \cite{basu}, 
where a perturbative approach was taken. In \cite{dias2} it was shown that
solitons can have arbitrarily large charge for large enough gauge coupling, 
while for small gauge coupling the solutions exhibit a spiraling behaviour towards
a critical solution with finite charge and mass. The stability of Reissner-Nordstr\"om-AdS (RNAdS)
solutions was also studied in this paper. It was found that for small gauge coupling RNAdS black holes
are never unstable to condensation of a massless, charged scalar field, while for 
intermediate gauge couplings RNAdS black holes become unstable for sufficiently large charge.
For large gauge coupling RNAdS black holes are unstable to formation of massless scalar hair for
all values of the charge. Moreover, it was observed that for large gauge coupling and
small charges the solutions exist all the way down to vanishing horizon. The limiting
solutions are the soliton solutions mentioned above. On the other hand for large charge the
limiting solution is a singular solution with vanishing temperature and finite entropy, which
is not a regular extremal black hole \cite{fiol1}. These results were extended to a tachyonic scalar 
field as well as to the rotating case \cite{brihaye_hartmannNEW}.
Recently, solutions in asymptotically global AdS in $(3+1)$ dimensions have been studied in \cite{menagerie}.
It was pointed out  that the solutions tend to their planar counterparts for large charges since 
in that case the solutions can become comparable in size to the AdS radius. The influence of the Gauss-Bonnet
corrections on the instability of these solutions has been discussed in \cite{Brihaye:2012cb}. 

In this paper, we are interested in Gauss-Bonnet solitons as well as black holes in $(4+1)$-dimensional
AdS space-time. We study the case of solitons in more detail and point out further details in comparison
to \cite{Brihaye:2012cb}. Moreover, we also discuss static and rotating Gauss-Bonnet black holes with and
without scalar hair. The rotating solutions without scalar hair have been previously studied in asymptotically
flat space-time \cite{Brihaye:2008kh,Brihaye:2010wx,Brihaye:2011hf} as well as in AdS space-time \cite{Brihaye:2008xu,Brihaye:2011hf}.
Our paper is organized as follows: we present the model in Section 2, while we discuss solitons in Section
3. In Section 4, we present our results for the black holes and we conclude in Section 5.

\section{The model}

In this paper, we are studying the formation of scalar hair on electrically charged black holes 
and solitons in $(4+1)$-dimensional Anti--de Sitter space--time. 
The action reads~:
\begin{equation}
S= \frac{1}{16\pi G} \int d^5 x \sqrt{-g} \left(R -2\Lambda + 
\frac{\alpha}{2}\left(R^{MNKL} R_{MNKL} - 4 R^{MN} R_{MN} + R^2\right) + 16\pi G {\cal L}_{\rm matter}\right) \ ,
\end{equation}
where $\Lambda=-6/L^2$ is the cosmological constant and $\alpha$  is the Gauss--Bonnet coupling with $0\leq \alpha \leq L^2/4$.
$\alpha=0$ corresponds to Einstein gravity, while $\alpha=L^2/4$ is the so-called Chern-Simons limit. 
${\cal L}_{\rm matter}$ denotes the matter Lagrangian which reads~:
\begin{equation}
{\cal L}_{\rm matter}= -\frac{1}{4} F_{MN} F^{MN} - 
\left(D_M\psi\right)^* D^M \psi - m^2 \psi^*\psi  \ \ , \ \  M,N=0,1,2,3,4  \ ,
\end{equation}
where $F_{MN} =\partial_M A_N - \partial_N A_M$ is the field strength tensor and
$D_M\psi=\partial_M \psi - ie A_M \psi$ is the covariant derivative.
$e$ and $m^2$ denote the electric charge and mass of the scalar field $\psi$, respectively.

The coupled gravity and matter field equations are obtained from the variation of the
action with respect to the matter and metric fields, respectively, and read
\begin{equation}
 G_{MN} + \Lambda g_{MN} + \frac{\alpha}{2} H_{MN}=8\pi G T_{MN} \ ,  \ M,N=0,1,2,3,4 \ ,
\end{equation}
where $H_{MN}$ is given by
\begin{equation}
 H_{MN}= 2\left(R_{MABC}R_N^{ABC} - 2 R_{MANB}R^{AB} - 2 R_{MA}R^{A}_N + R R_{MN}\right)
- \frac{1}{2} g_{MN} \left(R^2 - 4 R_{AB}R^{AB} + R_{ABCD} R^{ABCD}\right)
\end{equation}
and $T_{MN}$ is the energy-momentum tensor
\begin{equation}
 T_{MN}=g_{MN} {\cal L}_{\rm matter} - 2\frac{\partial {\cal L}_{\rm matter}}{\partial g^{MN}} \ .
\end{equation}

In the following, we want to study rotating Gauss-Bonnet solitons and black holes in $(4+1)$ dimensions. 
In general, such a solution
would possess two independent angular momenta associated to the two independent planes of 
rotation.
Here, we will restrict ourselves to the case where these two angular momenta are equal
to each other. The Ansatz for the metric reads \cite{kunz1}
\begin{eqnarray}
ds^2 & = & -b(r) dt^2 + \frac{1}{f(r)} dr^2 + g(r) d\theta^2 + h(r)\sin^2\theta \left(d\varphi_1 - 
\omega(r) dt\right)^2 + h(r) \cos^2\theta\left(d\varphi_2 -\omega(r)dt\right)^2 \nonumber \\
&+& 
\left(g(r)-h(r)\right) \sin^2\theta \cos^2\theta (d\varphi_1 - d\varphi_2)^2 \ ,
\end{eqnarray}
where $\theta$ runs from $0$ to $\pi/2$, while $\varphi_1$ and $\varphi_2$ are 
in the range $[0,2\pi]$.
The solution possesses two rotation planes at $\theta=0$ and $\theta=\pi/2$ and the isometry
group is $\mathbb{R}\times U(2)$.
In the following, we will additionally fix the residual gauge freedom by choosing $g(r)=r^2$. 

For the electromagnetic field and the scalar field we choose~:
\begin{equation}
A_{M}dx^M = \phi(r) dt  + A(r)\left(\sin^2\theta d\varphi_1 + \cos^2\theta d\varphi_2\right) \  \  
\  , \   \   \   \psi=\psi(r)  \ .
\end{equation}
Note that originally the scalar field is complex, but that we can gauge away the non-trivial
phase and choose the scalar field to be real.
The coupled, non-linear ordinary differential equations
depend on four independent constants: Newton's constant $G$, 
the cosmological constant $\Lambda$ (or Anti-de Sitter radius $L$), the charge $e$ and mass $m$ of the 
scalar field. Here, we set $m^2 = - 3/L^2$. The system possesses two scaling symmetries:
\begin{equation}
 r\rightarrow \lambda r \ \ \ , \ \ \ t\rightarrow \lambda t \ \ \ , \ \ \ L\rightarrow \lambda L
\ \ \ , \ \ \ e\rightarrow e/\lambda \ \ \ , \ \ \ A(r) \rightarrow \lambda A(r) \ \ \ ,  \ \ \ 
h(r) \rightarrow \lambda^2 h(r)  
\end{equation}
as well as 
\begin{equation}
 \phi \rightarrow \lambda \phi \ \ \ , \ \ \ \psi \rightarrow \lambda \psi \ \ \ , \ \ \ 
A(r) \rightarrow \lambda A(r) \ \ \ , \ \ \ 
e\rightarrow e/\lambda \ \ \ ,  \ \ \ \gamma\rightarrow \gamma/\lambda^2
\end{equation}
which we can use to set $L=1$ and $\gamma$ to some fixed value without loosing generality.
In the following, we will choose $\gamma = 9/40$ unless otherwise stated. 

Note that the metric on the AdS boundary in our model is that of a static 4-dimensional
Einstein universe with boundary metric $\gamma_{\mu\nu}$, $\mu$, $\nu=0,1,2,3$ given by
\begin{equation}
 \gamma_{\mu\nu}dx^{\mu} dx^{\nu} = -dt^2 + L^2 \left(d\theta^2 + \sin^2\theta d\varphi_1^2 + \cos^2\theta d\varphi_2^2\right)
\end{equation}
and is hence non-rotating. This is different to the case of rotating and charged black holes
in 4-dimensional AdS space-time, the so-called Kerr-Newman-AdS solutions \cite{carter1}, which
possess a boundary theory with non-vanishing angular velocity

Asymptotically, the matter fields behave as follows
 \be
     \phi (r\gg 1) = \mu + \frac{Q}{r^2} + \dots \ \ , \ \ A(r\gg 1)=\frac{Q_m}{r^2} + O(r^{-4}) \ \ , \ \ \psi(r\gg 1) =
 \frac{\psi_-}{r^{\lambda_-}} + \frac{\psi_+}{r^{\lambda_+}} + \dots
\ee
with 
\be
\lambda_{\pm} = 2 \pm \sqrt{4 + m^2 L_{eff}^2} \ \ , \ \ L_{eff}^2 = \frac{2 \alpha}{1 - \sqrt{1 - 4 \alpha/ L^2}} \ ,
\ee
where $Q$ and $Q_m$ are related to the electric and magnetic charge 
of the solution, respectively. $\mu$ is a constant that within the context of the gauge/gravity
duality can be interpreted as chemical potential of the boundary theory.  
For the rest of
the paper we will choose $\psi_{-}=0$. Within the AdS/CFT correspondence $\psi_+$ corresponds to
the expectation value of the boundary operator
in the dual theory.

The metric functions have the following asymptotic behaviour 
\begin{eqnarray}\
\label{bcads1}
 f(r >>1)&=&1+\frac{r^2}{L_{\rm eff}^2} + \frac{f_2}{r^2} +  O(r^{-4}) \ \ , \ 
\ b(r>>1)=1+\frac{r^2}{L_{\rm eff}^2} + \frac{b_2}{r^2} + O(r^{-4}) \ \ , \nonumber \\
h(r>>1)&=&r^2 + L_{\rm eff}^2 \frac{f_2-h_2}{r^2} + O(r^{-6}) \ \ , \ \  
\omega(r>>1)=\frac{\omega_4}{r^4} + O(r^{-8}) \ .
\end{eqnarray}

The parameters in the asymptotic expansion can be used to determine the mass and angular momentum of the
solutions. 
The energy $E$ and angular momentum $\bar{J}$ are 
\begin{equation}
 E=\frac{V_3}{8\pi G} 3 M \ \ {\rm with} \ \ M=\frac{f_2 - 4 b_2}{6} \ \ , \ \ 
\bar{J}=\frac{V_3}{8\pi G} J \ \ {\rm with} \ \ J=\omega_4   \ .
\end{equation}

In the following, we have constructed soliton as well as black hole solutions to the equations
of motions numerically using a Newton-Raphson method with adaptive grid scheme \cite{colsys}.

\section{Solitons}
In the following, we would like to study globally regular, i.e. soliton-like solutions to the equations
of motion. As was shown previously in \cite{brihaye_hartmannNEW} properly rotating solitons do not exist
in our model. We will hence concentrate on the static case which 
corresponds to the limit $\omega(r)\equiv 0$, $A(r)\equiv 0$ and $g(r)=h(r)=r^2$. 
In the static case, it is also convenient to work  with $a(r) \equiv \sqrt{b(r)/f(r)}$.
 For the remaining
functions, we have to fix appropriate boundary conditions which read
\be
  f(0) = 1 \ \ , \ \ \phi'(0) = 0  \ \ , \ \ \psi'(0) = 0
\ee
at the origin, where $\psi(0)\equiv\psi_0$ is a free parameter, and
\be
   a(r \to \infty) = 1 \ \ , \ \ \psi_- = 0
\ee 
on the conformal boundary. Note that the condition $\psi(0) = \psi_0$ 
can be replaced by fixing the electric charge $r^3 \phi'(r)|_{r \to \infty}= 2Q$. However, we found it
more convenient to fix $\psi_0$ and determine the charge $Q$ in dependence on this parameter.

Once fixing the parameters $e^2$ and $\alpha$ we can construct families of 
regular solutions labelled by the charge $Q$ or by the mass $M$. 
The pattern of solutions turns out to be so involved that neither $Q$ nor $M$ can be used to characterize the
solution uniquely. This is shown in Fig.\ref{fig_1}, where we give the charge $Q$ as function of $a(0)$ and the mass $M$ 
as function of $\psi(0)$, respectively. We observe that for small values of $e^2$ several disconnected
branches exist. These solutions have the same values of $M$ and $Q$, but differ in the values of
$\psi(0)$ and $a(0)$. We can also formulate this statement differently: for a fixed value of
$\psi(0)$ and $a(0)$ more than one solution exists. The different solutions are then characterized
by different values of the mass $M$ and charge $Q$.

%
%
%
\begin{figure}[h]
\begin{center}
\subfigure[Charge $Q$ as function of $a(0)$
]{\label{q_a0}\includegraphics[width=8cm]{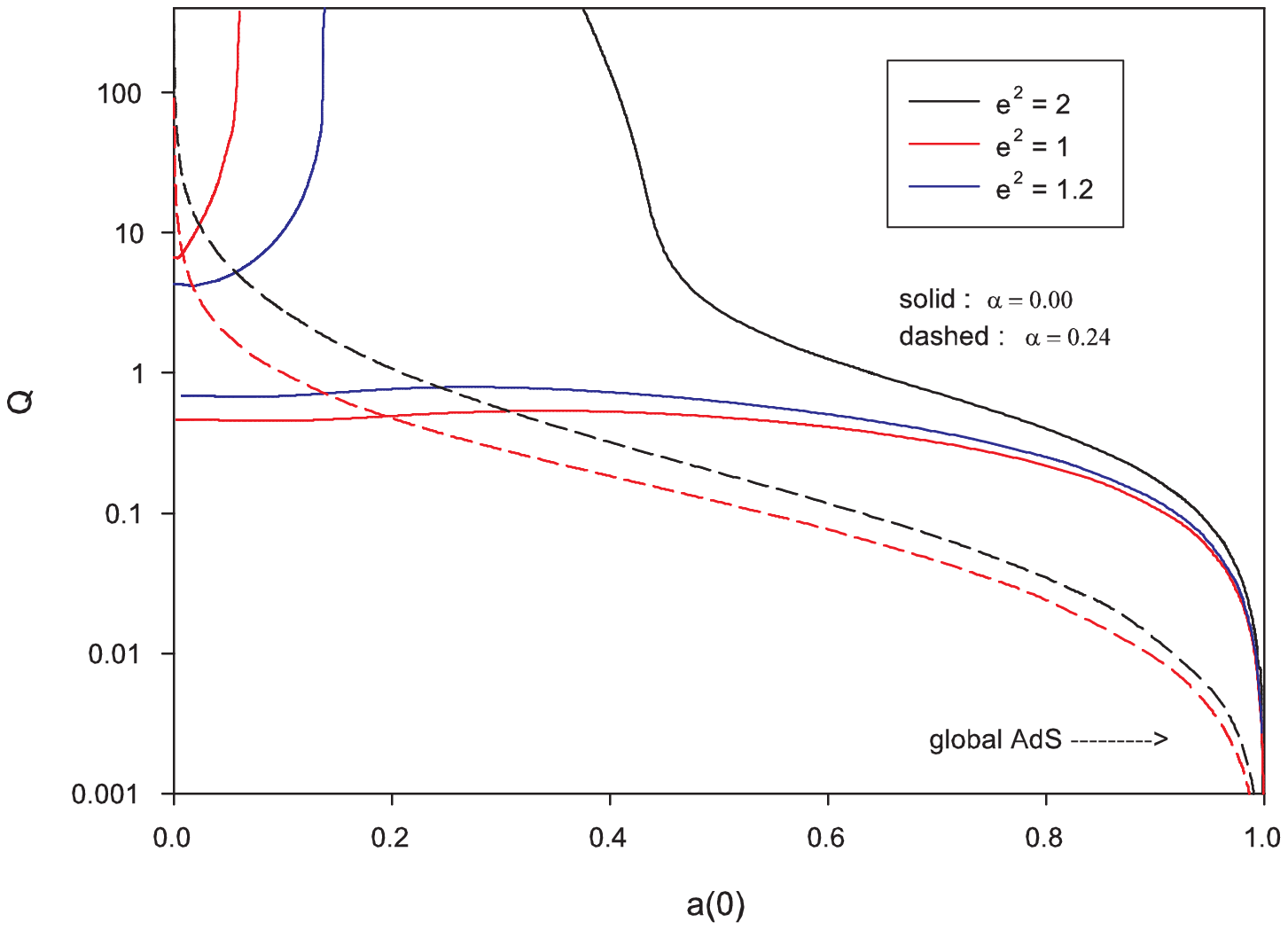}}
\subfigure[Mass $M$ as function of $\psi(0)$]
{\label{m_psi}\includegraphics[width=8cm]{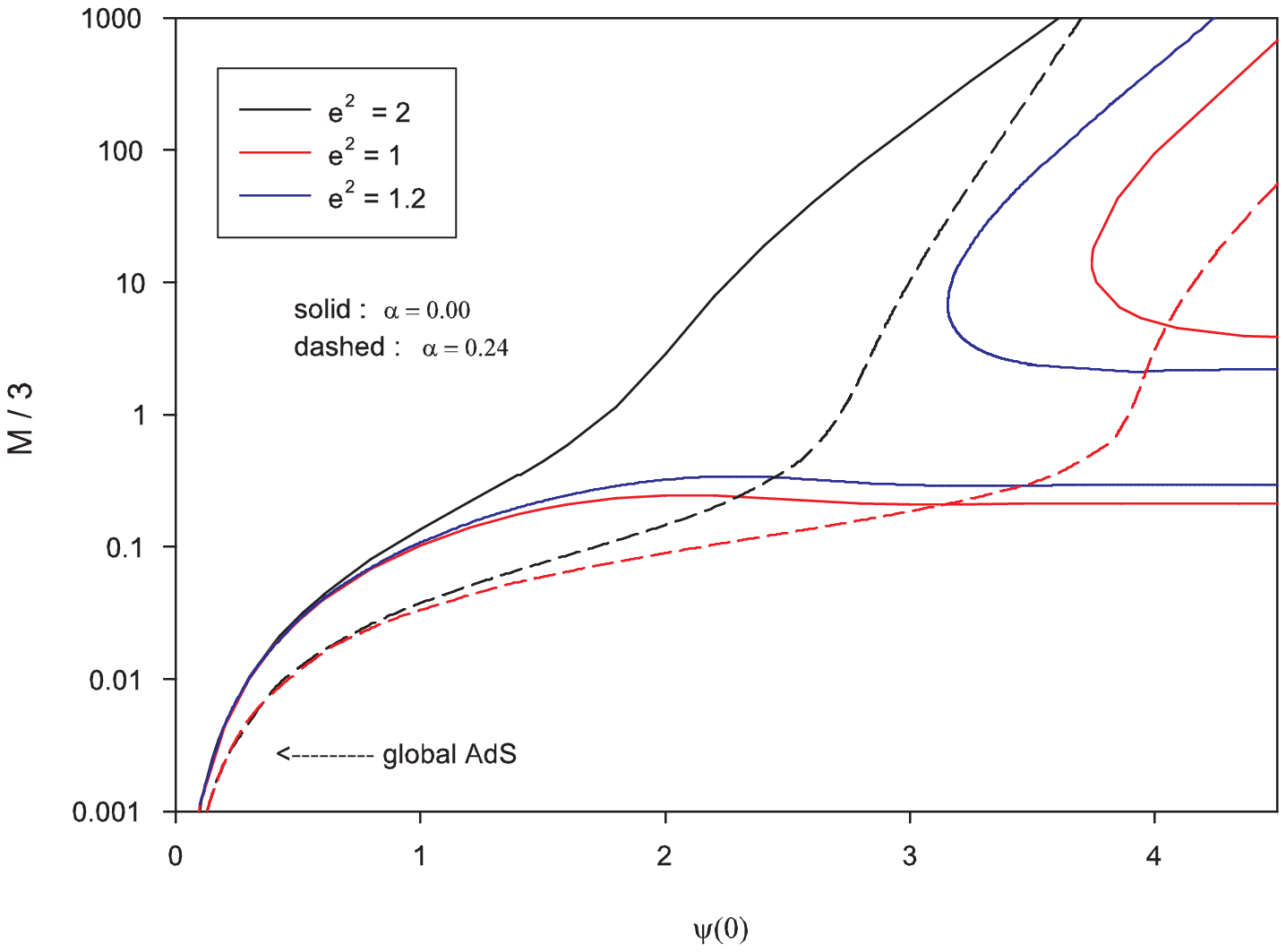}}
\end{center}
\caption{We show the charge $Q$ as function of $a(0)$ (left)
and the mass $M$ as function of $\psi(0)$ (right)
for three values of $e^2$ and two values of $\alpha$.
The limit corresponding to global AdS corresponds to $a(0) \rightarrow 1$ and $\psi(0) \rightarrow 0$. 
\label{fig_1}
}
\end{figure}
%
\begin{figure}[h]
\begin{center}
\subfigure[Fields  $f,a, \phi$
]{\label{q_a0}\includegraphics[width=8cm]{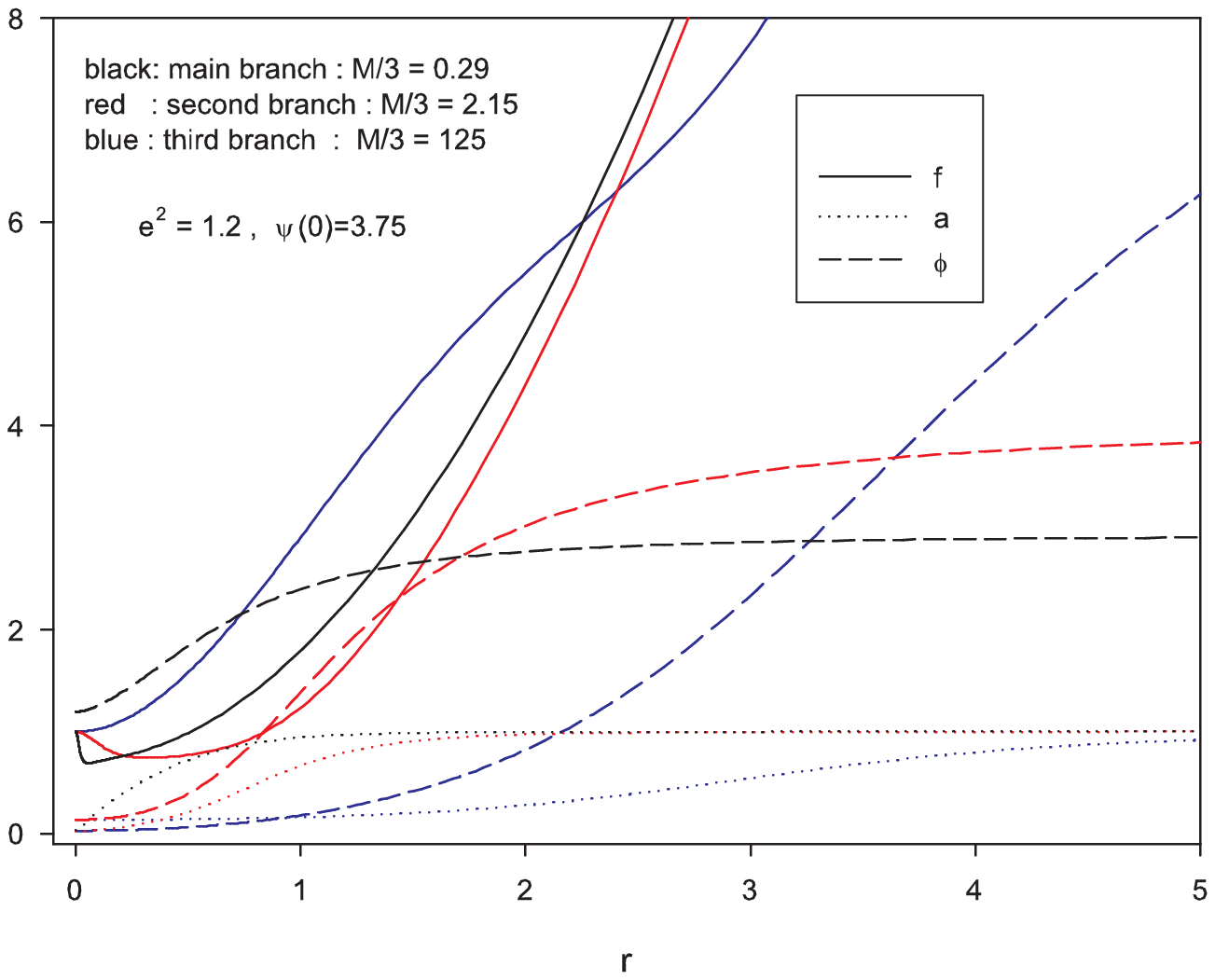}}
\subfigure[Fields $\phi', \psi, \psi'$]
{\label{m_psi}\includegraphics[width=8cm]{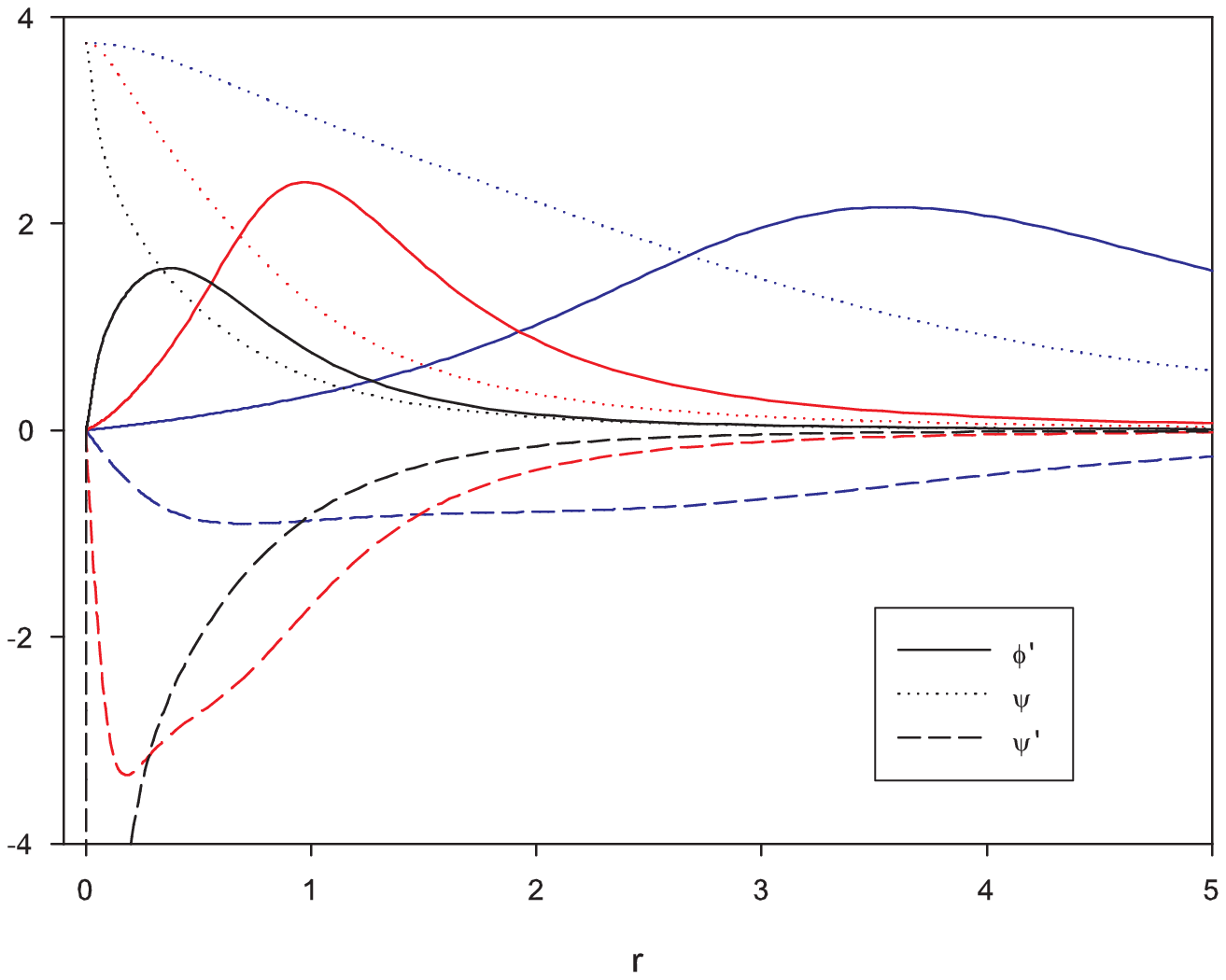}}
\end{center}
\caption{We show the metric functions $f$ and $a$, the electric potential $\phi$ (left) as well as
the scalar field function $\psi$ and the derivatives $\psi'$ and $\phi'$ (right)
of the three different soliton solutions that exist for $\alpha=0$, $e^2=1.2$ 
and $\psi(0)=3.75$.
\label{compare_3_e_12}
}
\end{figure}
\begin{figure}[h]
\begin{center}
\subfigure[Mass, charge and  $\phi(0)$ as function of $\alpha$]
{\label{q_a0}\includegraphics[width=8cm]{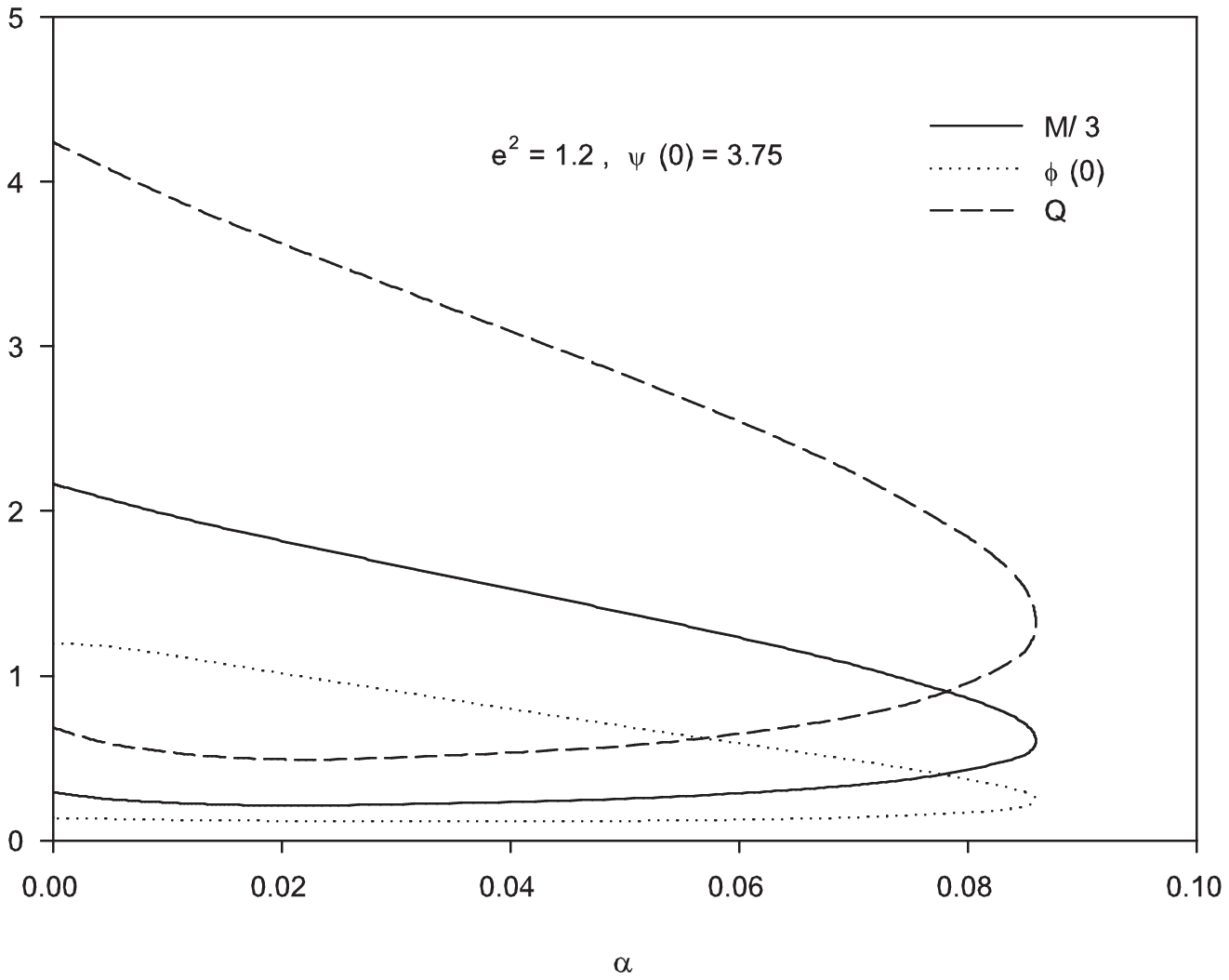}}
\subfigure[Mass, charge and  $\phi(0)$ as function of $\alpha$]
{\label{m_psi}\includegraphics[width=8cm]{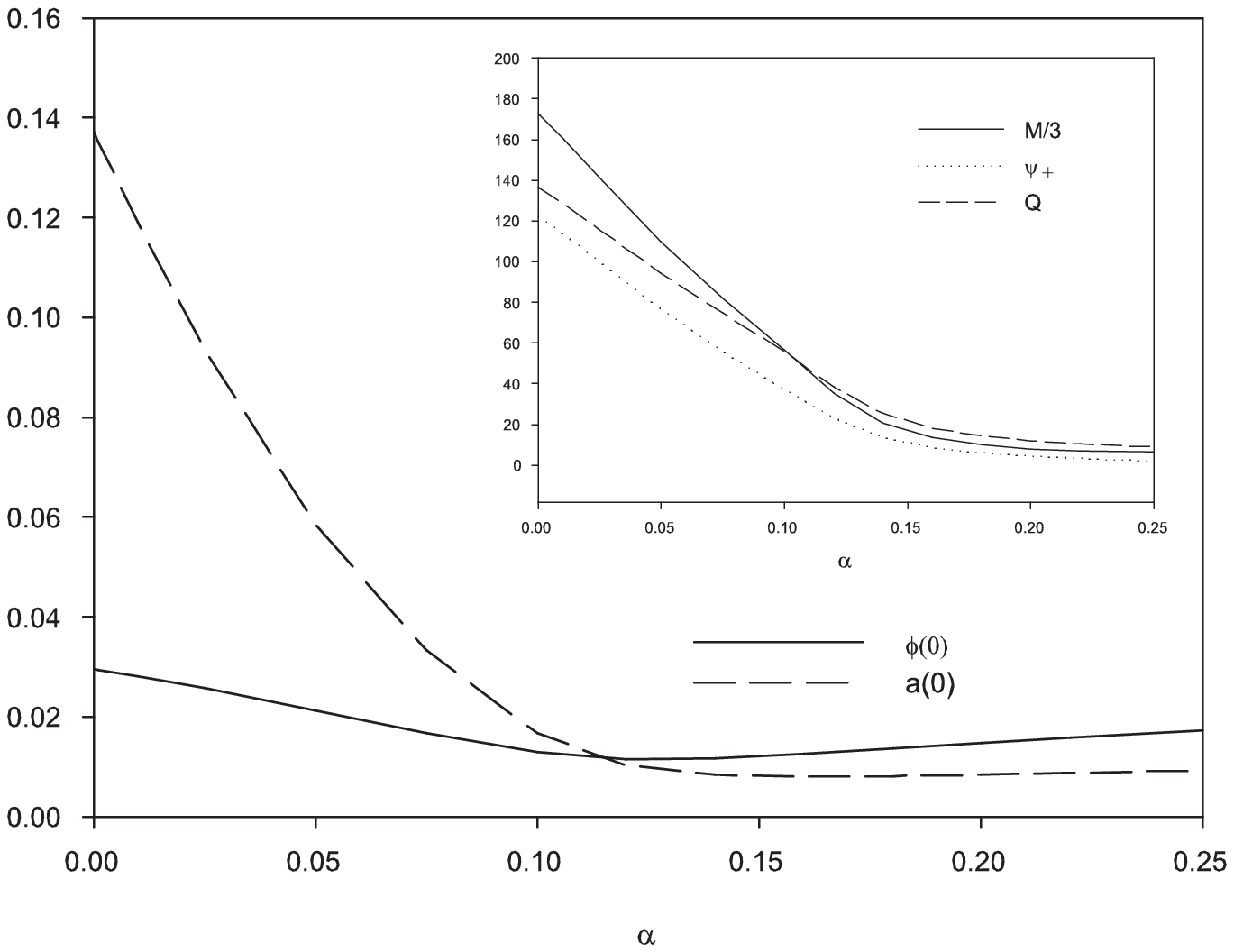}}
\end{center}
\caption{
We show the mass $M$, $\phi(0)$ as well as the charge $Q$ 
for the two lowest mass soliton solutions (left) as well as for the highest mass solution (right)
in dependence on $\alpha$. Here $\psi(0)=3.75$ and $e^2=1.2$.
\label{alpha_vary}
}
\end{figure}
\begin{figure}[h]
\begin{center}
\subfigure[Charge $Q$ as function of $a(0)$
]{\label{q_a0}\includegraphics[width=8cm]{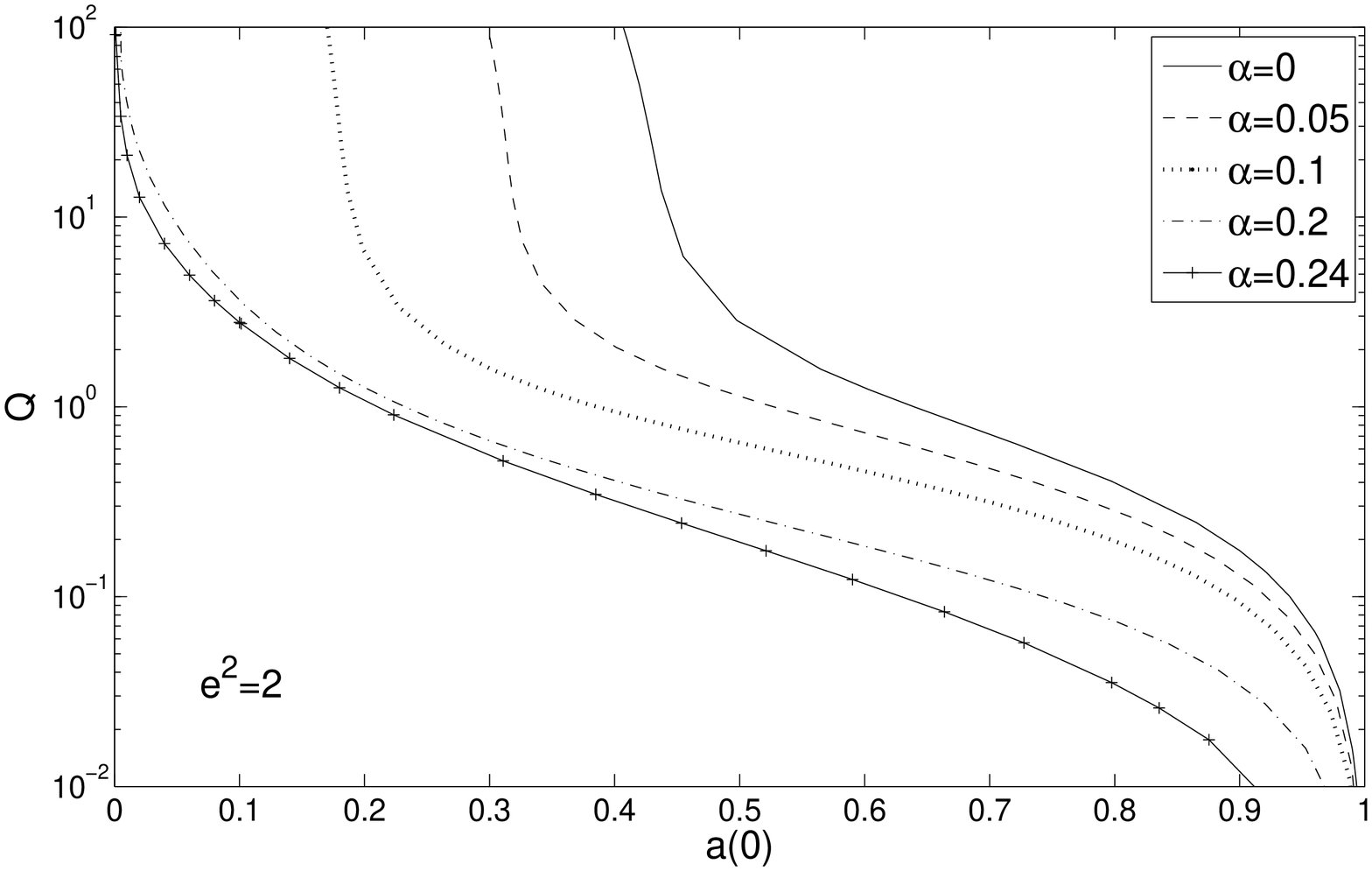}}
\subfigure[Mass $M$ as function of $\psi(0)$]
{\label{m_psi}\includegraphics[width=8cm]{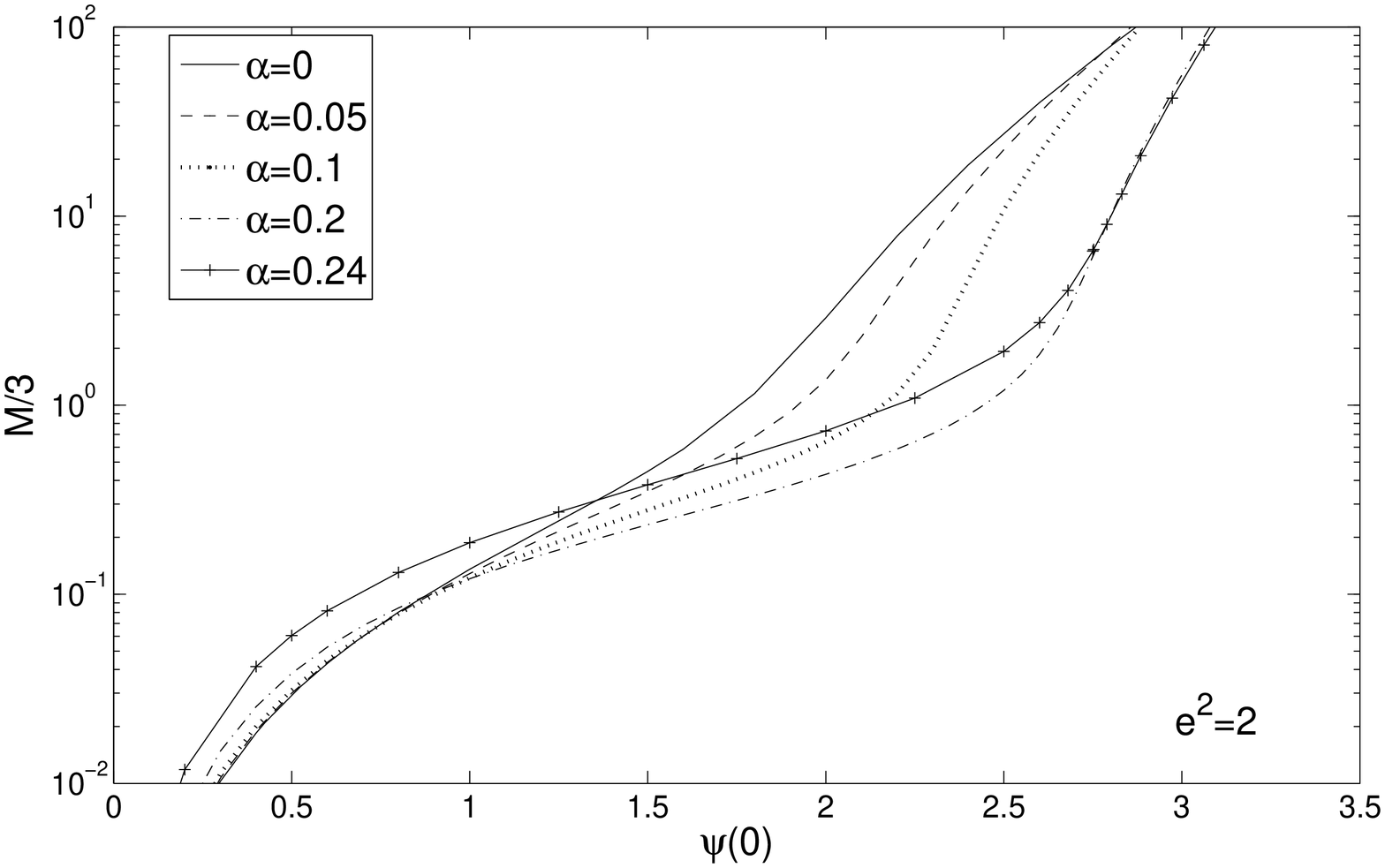}}
\end{center}
\caption{We show the charge $Q$ as function of $a(0)$ (left)
and the mass $M$ as function of $\psi(0)$ (right) 
for soliton solutions with $e^2=2$ and for different values of $\alpha$.\label{e22}}
\end{figure}

As an example let us consider the case $\alpha=0$, $e^2 = 1.2$. Here, we find 
that three soliton solutions exist with the same value of the scalar field at the 
origin $\psi(0)$. The profiles of these solutions are given in Fig. \ref{compare_3_e_12} (upper figure) for $\psi(0)=3.75$.
They have masses $M=0.29$, $M=2.15$ and $M=418.00$, respectively. 
The existence of several disconnected branches had been observed before in other models \cite{dias2,menagerie}.
Here, we find that the presence of the GB term changes the general pattern in
that at sufficiently large $\alpha > \alpha_{\rm cr}$ the two branches with the lowest mass join. 
This is shown in Fig.\ref{alpha_vary}
for $\psi(0)=3.75$ and $e^2=1.2$. For this choice of parameters, we find $\alpha_{\rm cr}\approx 0.086$.  
For $\alpha > \alpha_{\rm cr}$ these lowest mass solitons cease to exist. Contrary to that
the solution with the highest mass can be deformed all the way to the maximal value of $\alpha$, 
the Chern-Simons limit with $\alpha = L^2/4$. This is seen in Fig. \ref{alpha_vary}.

For larger values of $e^2$ we find that the qualitative pattern does not change when increasing $\alpha$.
This is shown in Fig.\ref{e22}, where we give the charge $Q$ in dependence on $a(0)$ and the mass $M$ in dependence
on $\psi(0)$ for $e^2=2$ and different values of $\alpha$. For $\alpha=0$ only one branch of soliton
solutions exists and this persists for $\alpha\neq 0$. No new branches appear due to the presence of the
GB term.

Let us mention here that the appearance of several branches is due to the presentation of data. If we would
instead consider to plot asymptotically measurable quantities as e.g. the mass $M$, the charge $Q$ or the free energy
$F=M-\mu Q$  we would find that
the solutions are uniquely described by these parameters. This is see in Fig.\ref{fig_new}, where we give the free energy
$F$ as function of the charge $Q$. 
The plot shows that for a given value of $Q$ there  is at most ONE solution and that the free energy decreases
with increasing $Q$.
For fixed $Q$ we find that the solution with the smallest $e^2$ has the lowest free energy.
If we fix $\psi(0)$ as described above up to three solutions exist of which we believe the lowest mass solution to be the
stable one. This solution is always on the same branch of solutions as the global AdS space-time with $M=Q=0$.

\begin{figure}
\centering
\epsfysize=8cm
\mbox{\epsffile{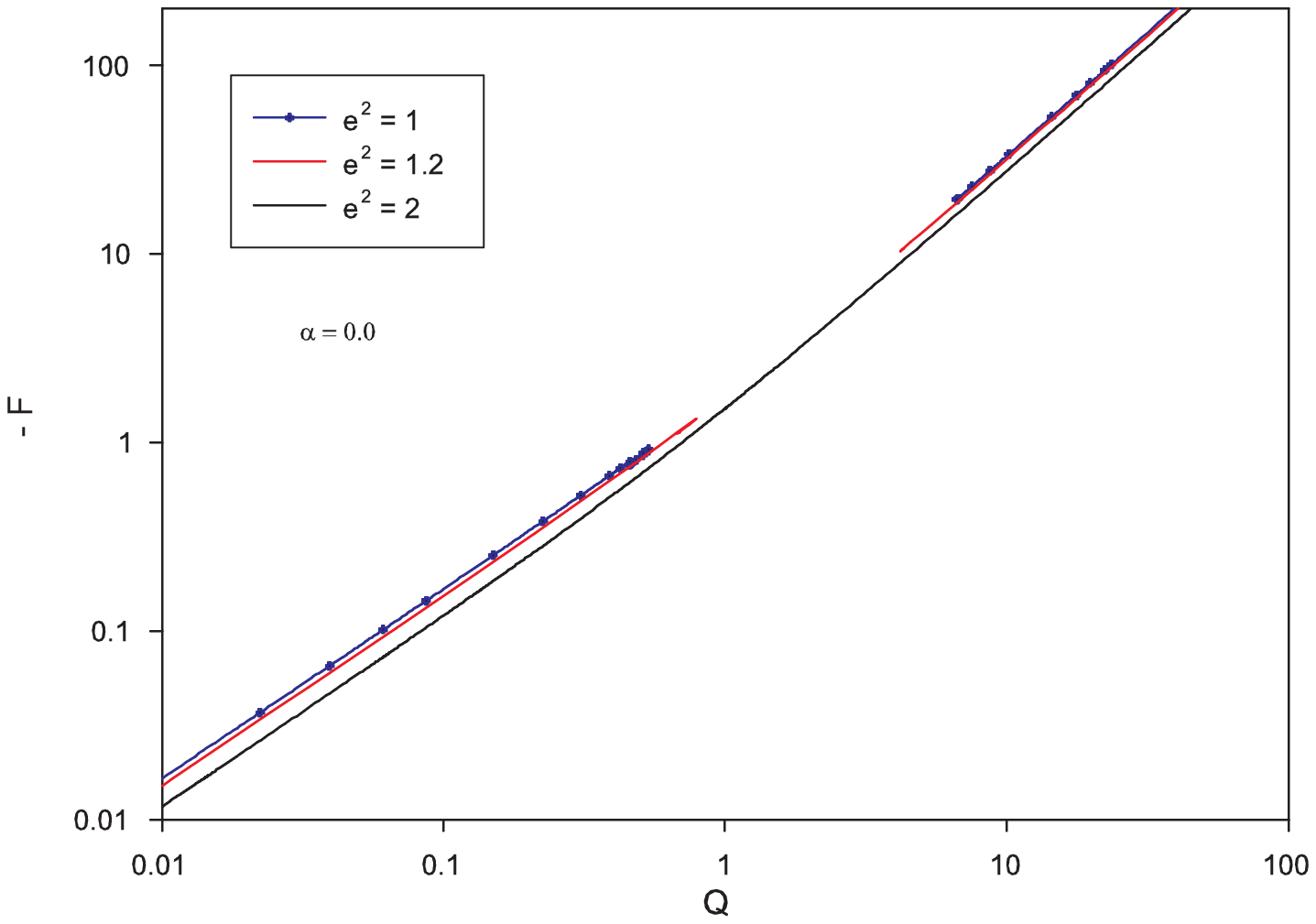}}
\caption{\label{fig_new}
We show the free energy $F=M-\mu Q$ of the soliton solutions in dependence on $Q$ for different
values of $e^2$ and $\alpha=0$. }
\end{figure}

Finally, we show the profiles of the metric and matter functions of the soliton solutions corresponding to large 
values of $\alpha$ in Fig.\ref{third_branch_alpha}. The solution with $\alpha=0.25$ corresponds to a hairy, charged
Chern-Simons soliton.

\begin{figure}
\centering
\epsfysize=8cm
\mbox{\epsffile{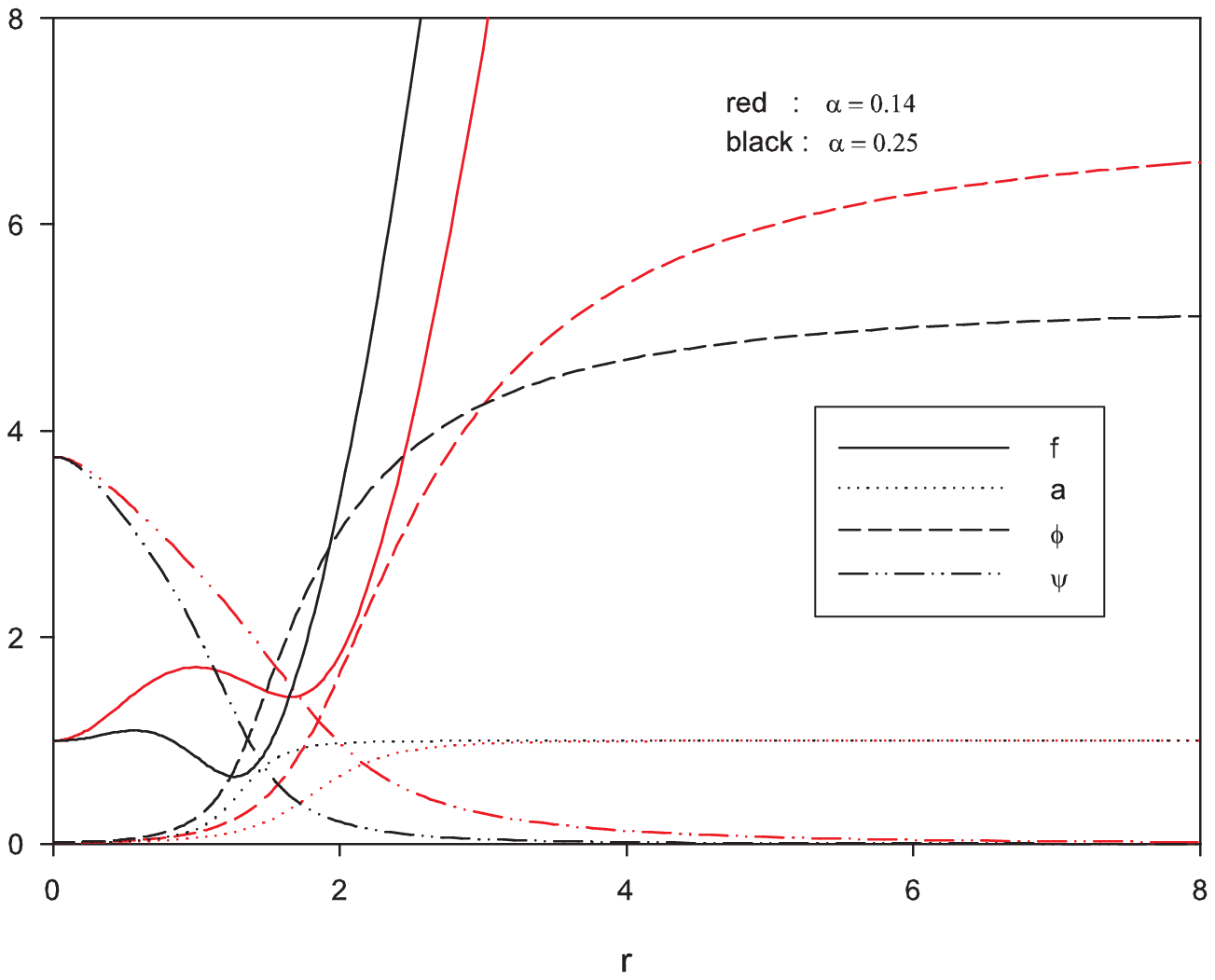}}
\caption{\label{third_branch_alpha}
We show the profiles of the metric functions $f$ and $a$ as well as of the electric potential $\phi$ and the
scalar field function $\psi$ of GB solitons for two values of  $\alpha$ and $\psi(0)=3.75$. }
\end{figure}

\section{Black holes}
We are interested in solutions possessing a regular horizon at $r=r_h$.
Hence we require
\be
\label{boundary_cond} 
f(r_h)=0 \ \ , \ \  b(r_h) = 0 \ \ , \ \ \ee
such that the metric fields have the following behaviour close to the horizon \cite{Brihaye:2008kh,
Brihaye:2010wx,Brihaye:2011hf,Brihaye:2008xu}
\begin{eqnarray}
 f(r)&=&f_1(r-r_h)+O(r-r_h)^2 \ \ \ , \ \ \ b(r)=b_1(r-r_h)+O(r-r_h)^2 \ \ \ , \nonumber \\
h(r)&=&h_0 + O(r-r_h) \ \ \ , \ \ \ \omega(r)=\omega(r_h) + w_1(r-r_h) + O(r-r_h)^2 \ \ , 
\end{eqnarray}
where $\omega(r_h)\equiv \Omega$ corresponds to the angular velocity at the horizon
and $f_1$, $b_1$ and $h_0$ are constants that have to be determined numerically. 
In addition there is a regularity condition for the metric fields on the horizon
given by $\Gamma_1(f,b,b',h,h',\omega,\omega')=0$, where $\Gamma_1$ is a lengthy polynomial
which we do not give here.
For the matter fields we have to require 
\begin{equation}
 \left. \left(\phi(r) + A(r) \omega(r)\right)\right\vert_{r=r_h}=0 \ , \  \Gamma_2(\psi, \psi')|_{r=r_h} = 0 \ ,
\end{equation}
where $\Gamma_2$ is a polynomial expression in the fields which we also do not present here.

Using the expansions of the metric functions we find that 
temperature $T_{\rm H}$ and the entropy $S$ are given by
\cite{Brihaye:2008kh}
\begin{equation}
 T_{\rm H}=\frac{\sqrt{f_1 b_1}}{4\pi} \ \ , \ \ S=\frac{V_3}{4G} r_h^2 \sqrt{h_0} \ , 
\end{equation}
where $V_3=2\pi^2$ denotes the area of the three-dimensional sphere with unit radius.

\subsection{Static black holes}
We first study the static case with $\Omega=0$. This has already been considered in \cite{Brihaye:2012cb}. Here we
point out further important details in the pattern of solutions.

\subsubsection{Exact solutions}
In the case $\psi\equiv 0$, i.e. when the scalar field vanishes there exists an explicit solution
to the equations. This reads
\cite{deser,Cai:2001dz,Cvetic:2001bk,Cho:2002hq,Neupane:2002bf}
\begin{equation}
      f(r) = 1 + \frac{r^2}{2\alpha} \left(1\mp\sqrt{1-\frac{4\alpha}{L^2} + 
\frac{4\alpha M}{r^4} - \frac{4\alpha\gamma Q^2}{r^6}}\right)  \ \ , \ \ a(r)=1 \ \ , \ \ 
      \phi(r) = \mu - \frac{Q}{r^2}  \ ,
\label{rn}
\end{equation}
where $M$ and $Q$ are arbitrary integration constants that can be interpreted as the mass and
the charge of the solution, respectively. Note first that for $M=Q=0$ there are two global AdS solutions with effective AdS radius
$L_{\rm eff}^2 = L^2/2\left(1\pm \sqrt{1-4\alpha/L^2}\right)$. 

Since we are interested in black hole solutions here
which fulfill $f(r_h)=0$ the ``+'' solution is of no interest in the static case.
In the limit $\alpha\rightarrow 0$, 
the metric function $f(r)$ of the ``-'' solution becomes $f(r)=1 +\frac{r^2}{L^2}-\frac{M}{r^2}+\frac{\gamma Q^2}{r^4}$
and the corresponding solutions are Reissner-Nordstr\"om-Anti-de Sitter (RNAdS) black holes.

\subsubsection{The $\alpha=0$ limit}
We first discuss the limit of vanishing Gauss-Bonnet coupling $\alpha=0$. 
We find that black hole solutions exist for generic values of $e^2$, $r_h$ and $\psi(r_h)$ but that
the domain of existence for these parameters is limited. It depends crucially on the value of $e^2$.
For large values of $e^2$ for which only a single branch of corresponding soliton 
solutions exist black holes exist for all values of $r_h > 0$ and $\psi(r_h) >0$. In the limit $\psi(r_h) \to 0$
with $r_h$ fixed 
they approach the RNAdS solution, while 
in the limit $r_h \to 0$ with $\psi(r_h)$ fixed they approach the corresponding soliton.

The situation is more subtle for small values of $e^2$, i.e. when disconnected branches of solitons exist. This is 
shown in Fig. \ref{fig_bh_4} for $e^2 = 1.2$, where we present the temperature and the mass of the solutions
for several values of $\psi(r_h)$. 
\begin{figure}[h]
\begin{center}
\subfigure[$T_H$ and $a(r_h)$ as function of  $r_h$
]{\label{q_a0}\includegraphics[width=8cm]{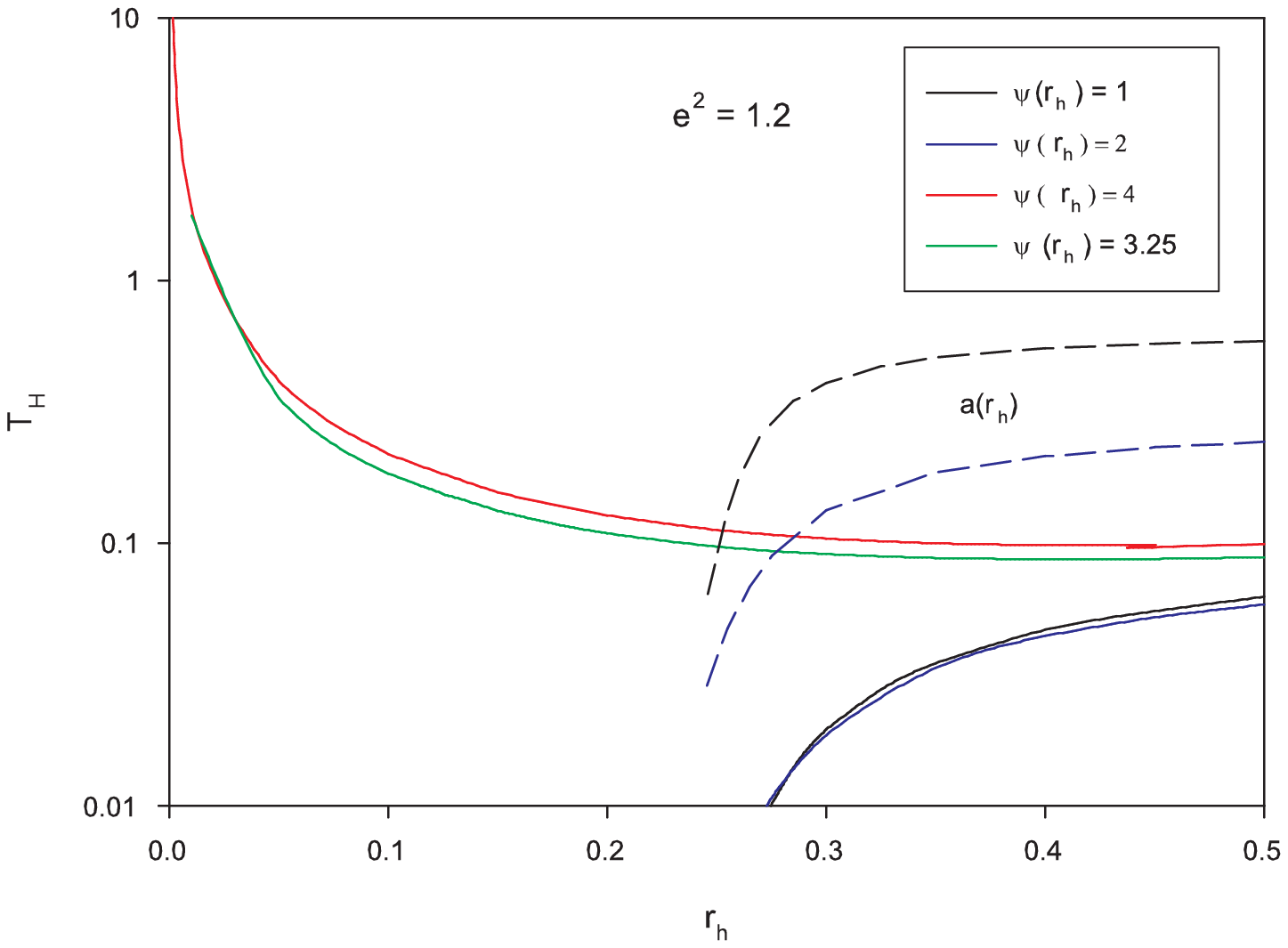}}
\subfigure[$M$ as function of   $r_h$]
{\label{m_psi}\includegraphics[width=8cm]{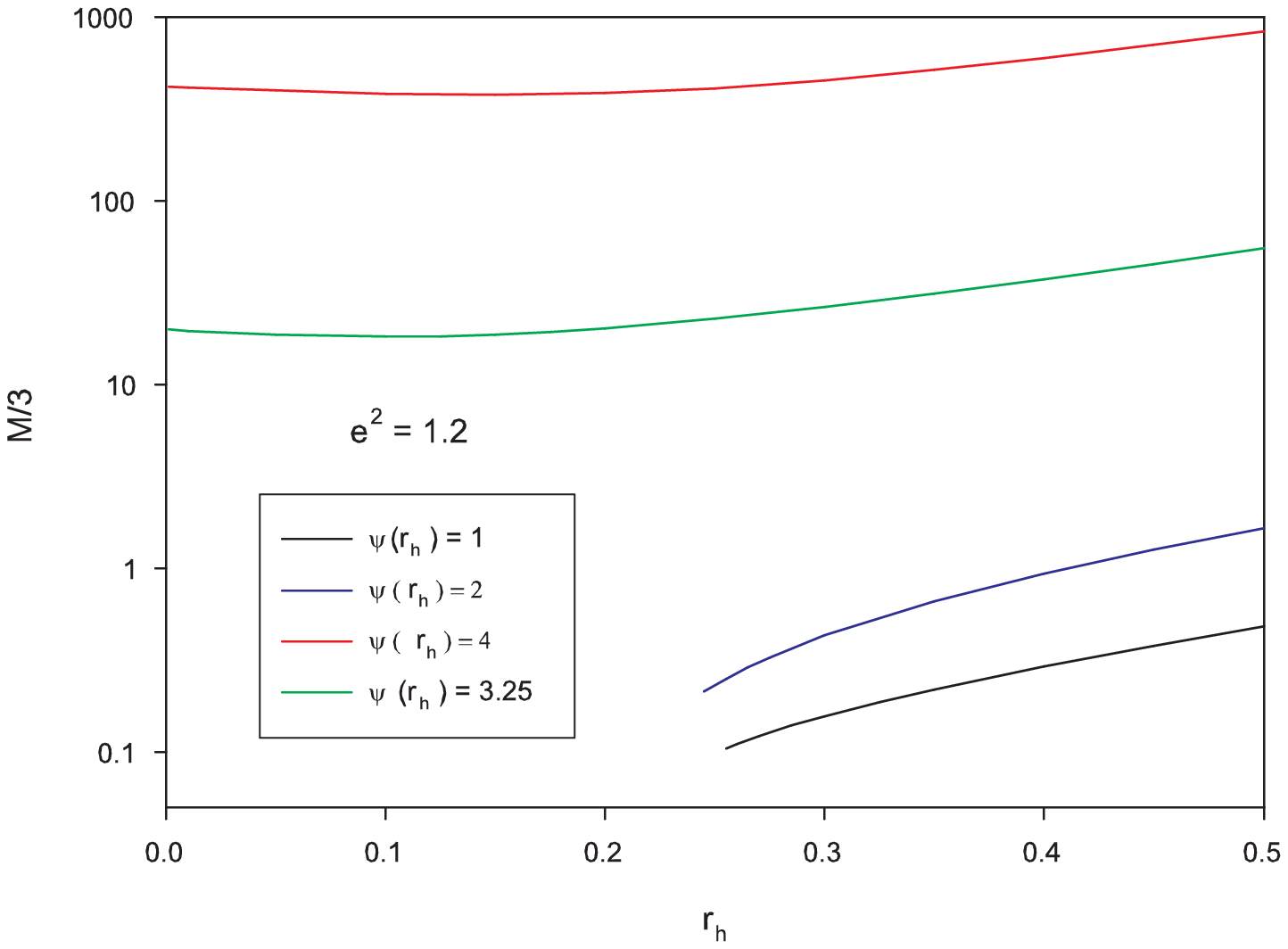}}
\end{center}
\caption{
We show the temperature $T_H$, the value $a(r_h)$ (left) and the mass $M$ (right) 
of black hole solutions with scalar hair  for several values of $\psi(r_h)$. 
Here $\alpha=0$ and $e^2=1.2$.
\label{fig_bh_4}}
\end{figure}
\begin{figure}[h]
\begin{center}
\subfigure[Mass and $T_H$ as function of  $\psi(r_h)$
]{\label{q_a0}\includegraphics[width=8cm]{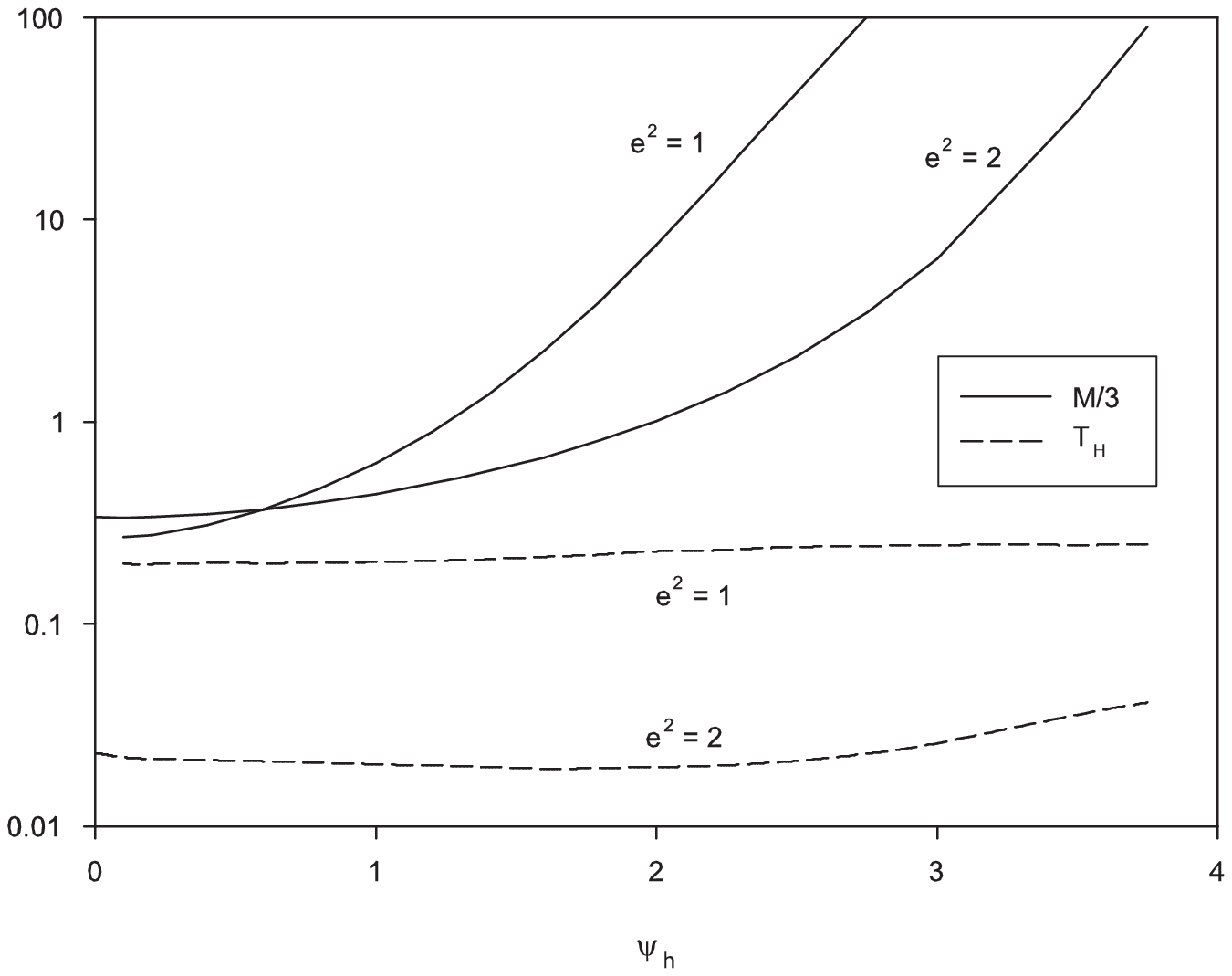}}
\subfigure[$M$ as function of   $Q$]
{\label{m_psi}\includegraphics[width=8cm]{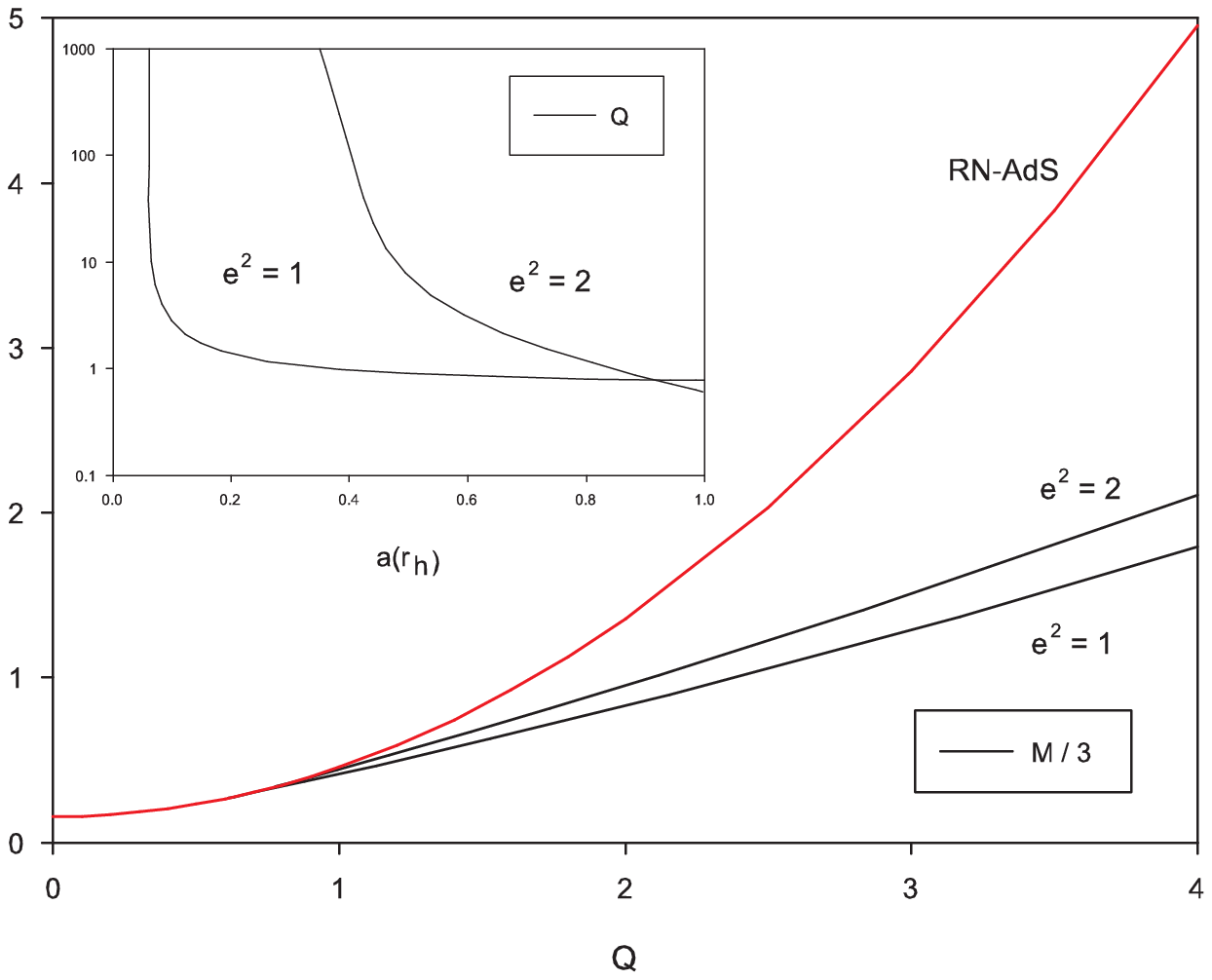}}
\end{center}
\caption{
We show the temperature $T_{\rm H}$ and the mass $M$ as function of $\psi_h\equiv \psi(r_h)$ (left)
as well as the mass as function of  charge $Q$ 
for RN (red line) and hairy (black lines) black holes for two different values of $e^2$ (right). 
Here $\alpha=0$ and $r_h=0.5$.
The charge as function of $a(r_h)$ is given in the insert. 
\label{fig_bh_2}}
\end{figure}
We find that for sufficiently high values of $\psi(r_h)$ the black holes exist all the way down to
$r_h=0$, where they join the branch of soliton solutions. In this limit the temperature goes to infinity and
the mass tends to the corresponding mass of the soliton solution. This is clearly visible for $\psi(r_h)=3.25$ and
$\psi(r_h)=4$, respectively. On the other hand, for small values of $\psi(r_h)$ we find that the black holes
exist only above a critical value of the horizon radius $r_h=r_h^{(cr)}$. For $r_h\rightarrow r_h^{(cr)}$ 
the temperature of the black hole tends to zero. However, this is not an extremal black hole, but a singular
black hole solution. This is apparent when observing that $a(r_h)\rightarrow 0$ in this limit. This is clearly
seen for $\psi(r_h)=1$ and $\psi(r_h)=2$ in Fig.\ref{fig_bh_4}. 

There should hence be a critical value of $\psi(r_h)$ at which the transition between the two pattern 
appears. We find that the determination of the exact numerical value of this critical $\psi(r_h)$ is quite involved, 
but we conjecture that it corresponds to the minimal value of $\psi(0)$ at which the disconnected branches
of solitons terminate. 

We further observe that if $r_h$ is big enough (typically $r_h\gtrsim 0.5$) 
a connected branch of black holes with
scalar hair labelled by $\psi(r_h)$ exists. This is shown in  
Fig. \ref{fig_bh_2} for $e^2=1$ and $e^2=2$, respectively.
In this figure we show the mass and temperature as function of $\psi_h\equiv \psi(r_h)$ 
as well as the charge as function of $a(r_h)$  of these black hole solutions.
For $\psi(r_h) \to 0$, $a(r_h)\to 1$ these solutions approach the RNAdS solutions with finite values of the
mass $M$, charge $Q$ and temperature $T_{\rm H}$.

\subsubsection{GB black holes}
In the following, we restrict our analysis to a finite number of parameters and construct
mostly (unless otherwise stated) black holes with $r_h=0.5$ and $e^2=1$ or $e^2=2$. We can then
discuss the pattern of solutions for different values of $\alpha$.  

We find that for $\alpha\neq 0$ and for some values of the parameters $e^2$ and $\psi(r_h)$,
the black holes can be continued up to the Chern-Simons limit $\alpha=L^2/4$. 
This is the case e.g. for $e^2=1$, $\psi(r_h)=3.75$, $r_h=0.5$
as shown in Fig. \ref{fig_bh_3}. For larger values of $e^2$ (with the same values of $\psi(r_h)$ and $r_h$) 
the black hole ceases to exist at
some intermediate value of $\alpha$ where the temperature $T_{\rm H}$ tends to zero. This is shown in Fig.\ref{fig_bh_3}
for $e^2=2$. To state it differently, for sufficiently large values of $e^2$ GB black hole
solutions with scalar hair exist only up to a critical value of the GB coupling that is smaller
than the Chern-Simons limit. For our specific choice of parameters here we find that solutions exist only for
$\alpha \lesssim 0.14$. 

\begin{figure}
\centering
\epsfysize=8cm
\mbox{\epsffile{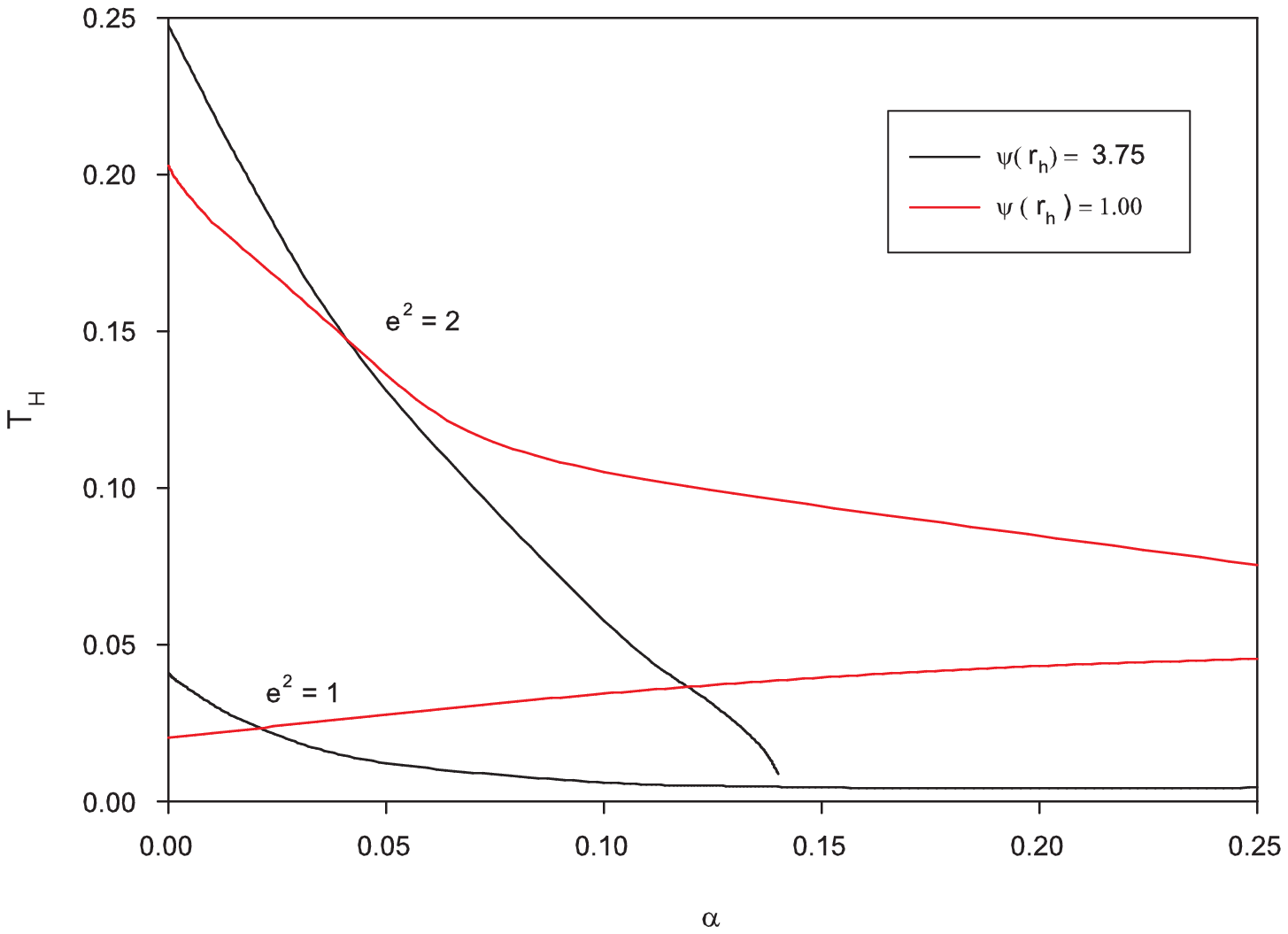}}
\caption{\label{fig_bh_3}
We show the temperature $T_H$ of the GB black holes with scalar hair 
as function of  $\alpha$ for $e^2=1$ and $e^2=2$, respectively. 
Here $\psi(r_h)=3.75$ and $r_h=0.5$.}
\end{figure}

\subsection{Rotating black holes}
In principle, we want to discuss rotating black holes with scalar hair in this paper. However, during our numerical
analysis it turned out that some as yet unnoticed features appear also for rotating GB black holes without scalar
hair. This is why we will discuss this case first before turning to the case with scalar hair. 

\subsubsection{Black holes without scalar hair}
In order to understand the pattern of solutions for $\psi\equiv 0$, it is useful to recall some limiting cases.
Non-rotating solutions with $\Omega=0$  can be constructed for $\alpha \in [0, L^2/4]$, $Q \in [0,Q_m]$, where $Q_m$ is
the maximal charge up to where the solutions exist. 
For $\alpha=0$ the solutions with $Q$ fixed exist up to a maximal value of the
rotation parameter $\Omega = \Omega_m$ where they become extremal \cite{Brihaye:2011hf}.
For $Q=0$ the solutions with a fixed $\alpha$ also exist up a maximal value of $\Omega$
as discussed for $L^2 = \infty$ in \cite{Brihaye:2010wx}  and for $L^2 < \infty$ in \cite{Brihaye:2011hf} .  

Let us first discuss how the solutions evolve when gradually increasing the horizon velocity
$\Omega$. Our numerical results
indicate that a branch of solutions 
can be constructed up to a critical value $\alpha=\alpha_{cr}$ which depends on $\Omega$ such that these solutions exist for
\be
                 0 \leq \alpha < \alpha_{\rm cr}(\Omega) \ \ , \ \ \alpha_{\rm cr}(\Omega) 
< \alpha_{\rm cr}(0) = \frac{L^2}{4}  \ .
\ee 
The critical value of $\alpha=\alpha_{\rm cr}$ up to where rotating GB black holes exist 
depends  both on $\Omega$ and $Q$. We find e.g. for small values of $\Omega$,
$r_h=0.8$, $Q=1$ that $\alpha_{\rm cr} \approx 0.23$. 

The  critical phenomenon occurring in the limit $\alpha \to \alpha_{\rm cr}$ can be understood by examining the
value $h'(r_h)$. It turns out that in this limit, the value $h'(r_h)$ (which is  a positive number 
for $\alpha$ large enough)  increases very rapidly with
$\alpha$. Our numerical results further suggest that 
this branch cannot be continued for  $\alpha > \alpha_{\rm cr}$. This is shown in Fig.\ref{q_vary}, 
where the parameter $h'(r_h)$ is given as a function of $\alpha$ for different values of $Q$ and $\Omega$.
\begin{figure}
\centering
\epsfysize=8cm
\mbox{\epsffile{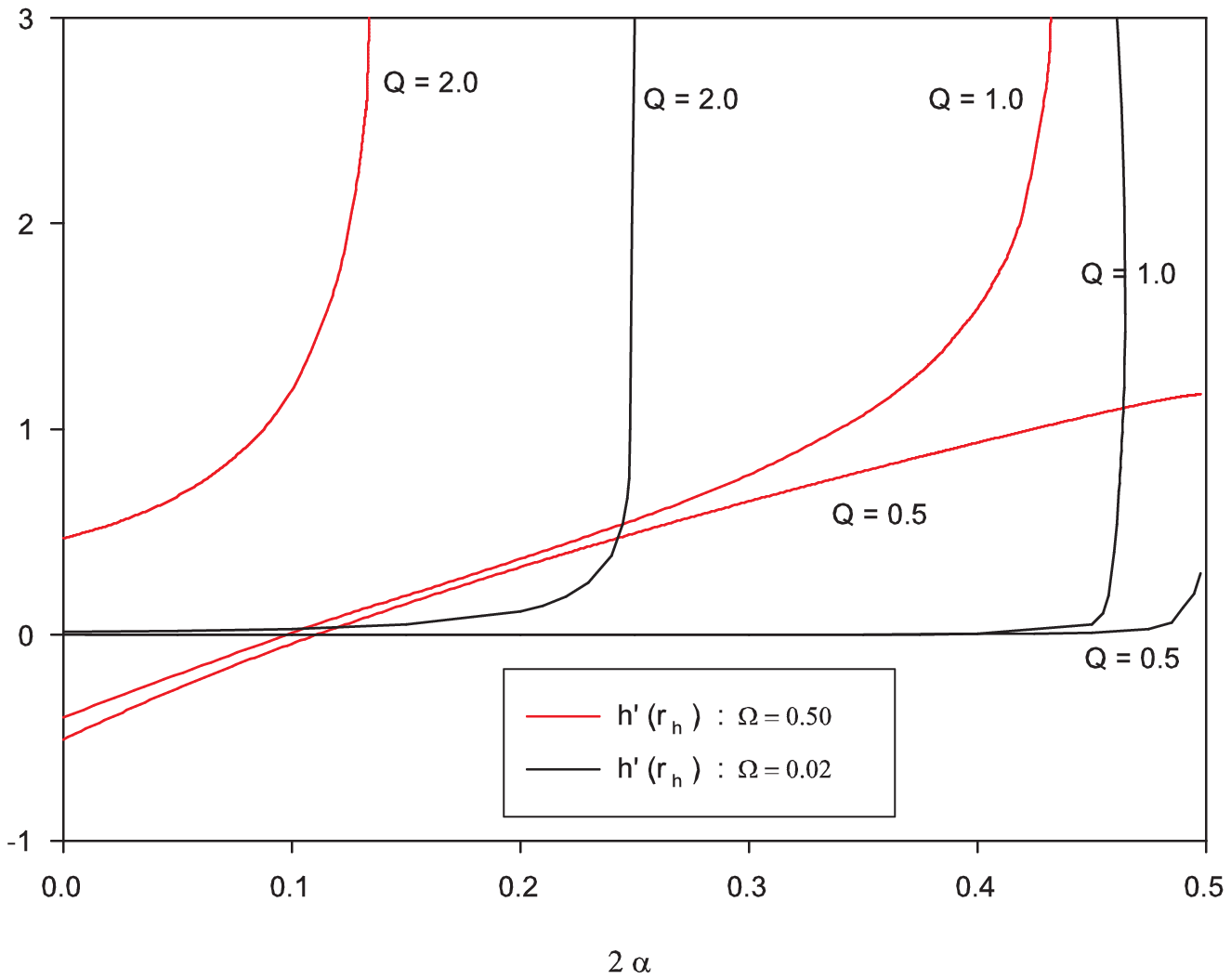}}
\caption{\label{q_vary}
We show the value $h'(r_h)$ as function of $\alpha$ for $\Omega = 0.02, 0.5$ and $Q=0.5,1.0,2.0$. 
Here  $r_h=0.8$, $\psi\equiv 0$. }
\end{figure}
\begin{figure}
\centering
\epsfysize=8cm
\mbox{\epsffile{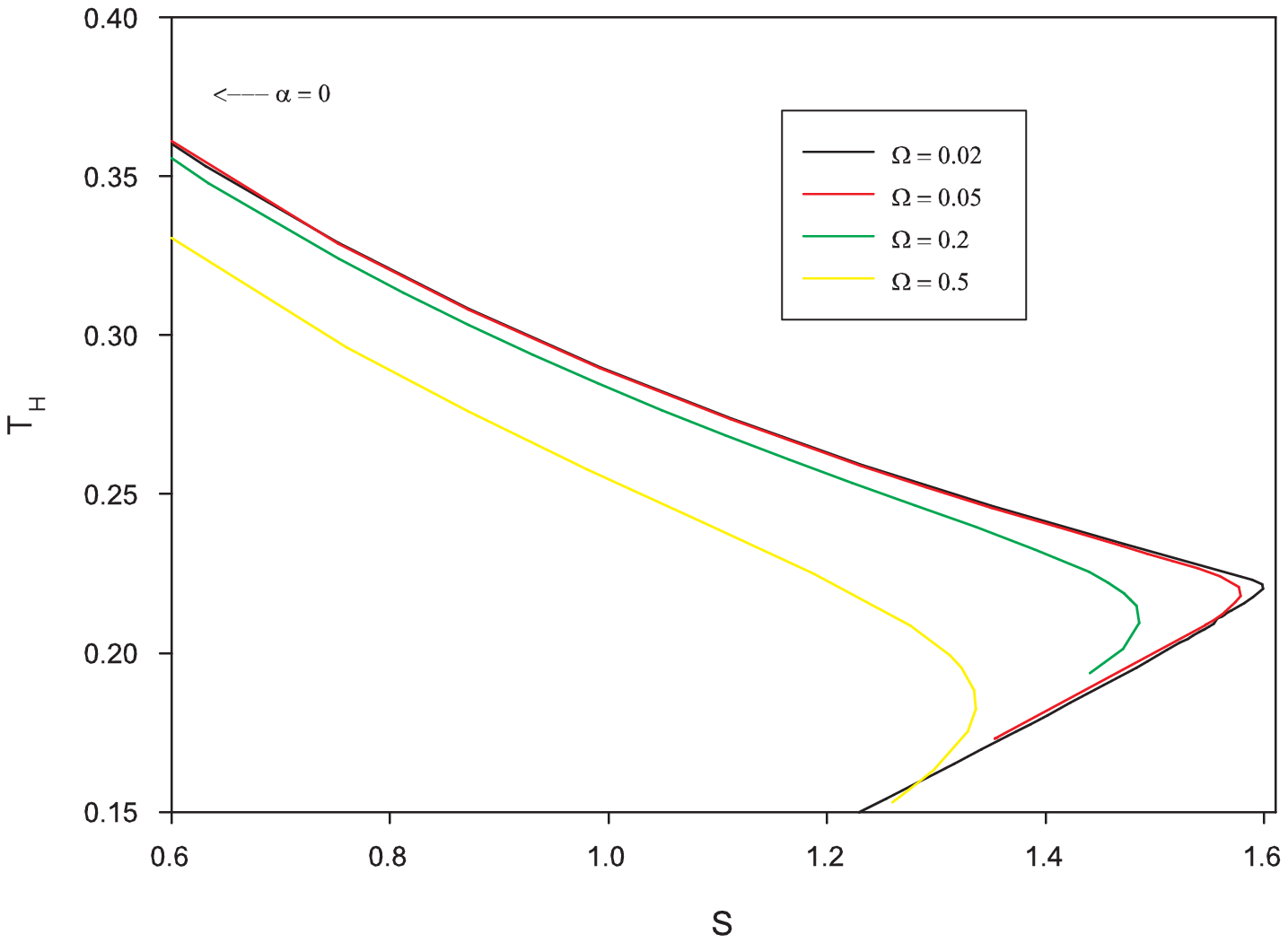}}
\caption{\label{new_alpha_vary_TS}
We show the temperature $T_{\rm H}$ as function of the entropy $S$ for  $\Omega = 0.02$, $0.05$, $0.2$, $0.5$. 
Here $Q=1$, $r_h=0.8$ 
and $\psi\equiv 0$.  Note that small (large) entropy and large (small) temperature
corresponds to $\alpha$ small (large).
}
\end{figure}

In order to understand the physical meaning of this pattern
we present 
the temperature $T_{\rm H}$ as function of entropy $S$ in Fig.\ref{new_alpha_vary_TS}. 
This clearly shows that two branches with a transition between the branches occurring at an 
intermediate value of $\alpha$ exist. Note that small (large) entropy and large (small) temperature
corresponds to $\alpha$ small (large). We observe that for sufficiently large $\alpha$ the entropy increases
with increasing $T_{\rm H}$ which would signal thermodynamical stability.

Increasing the value of $h'(r_h)$ further we were able to construct
a second branch of solutions for $\alpha < \alpha_{\rm cr}$. The two branches are such that
they merge at $\alpha = \alpha_{\rm cr}$. This is shown in Fig. \ref{new_alpha_vary}
where we present some physical quantities like the mass $M$ and the temperature $T_{\rm H}$ 
in dependence on $\alpha$
for a rotating GB black hole solution with $\Omega=0.02$, $Q=1$ and $r_h=0.8$. 
Although the numerical construction becomes quite involved we strongly suspect that
further branches can be constructed in the region around $\alpha_{\rm cr}$.  
These branches, however, cannot be extended 
to small values of $\alpha$. Surprisingly, we find that 
for the values of $\alpha$ for which the two solutions coexist
the solutions of the second branch  have  smaller energy than the solutions
on the first branch. This is demonstrated in  Fig.\ref{new_alpha_vary}.

\begin{figure}
\centering
\epsfysize=8cm
\mbox{\epsffile{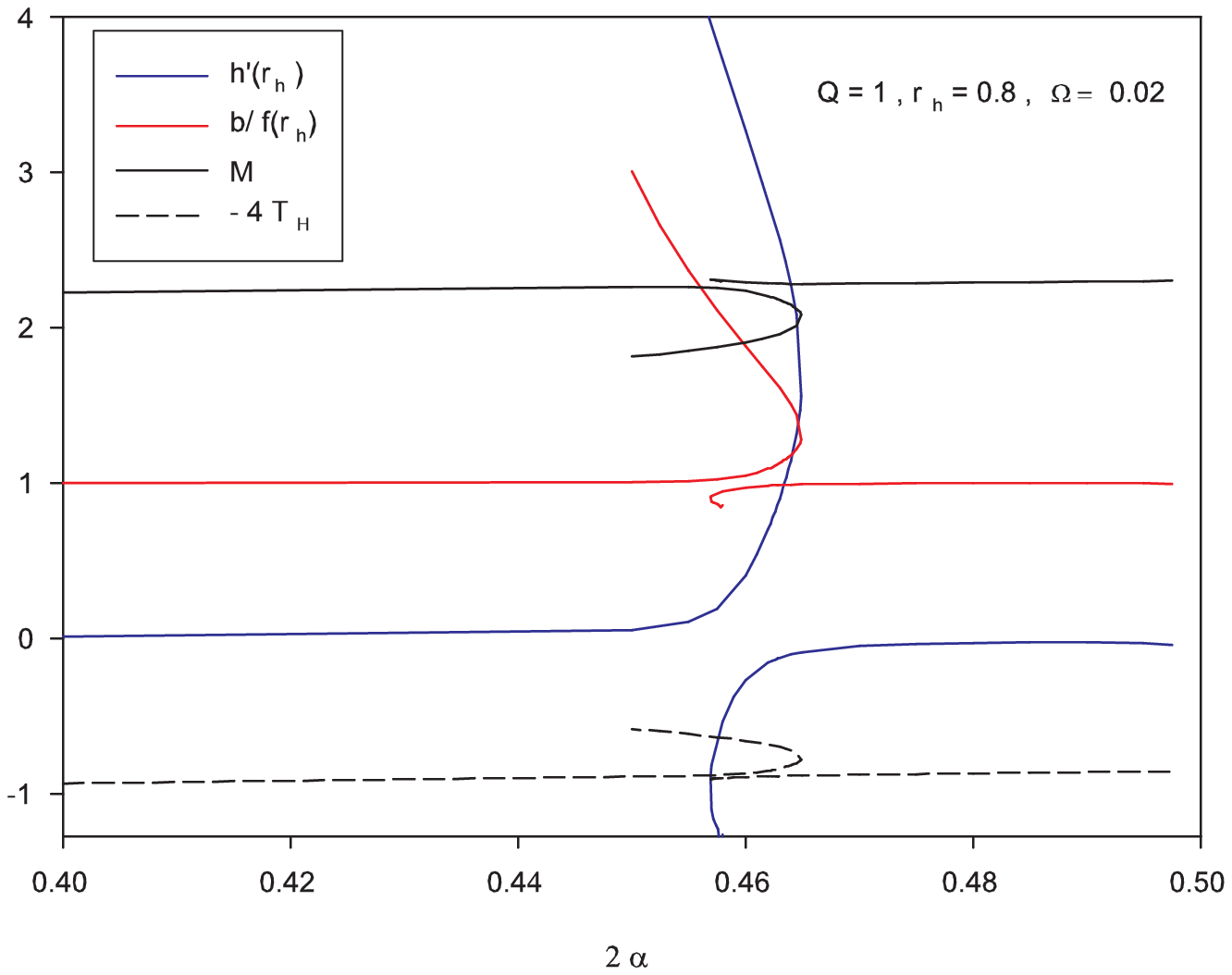}}
\caption{\label{new_alpha_vary}
We show the mass $M$, the temperature $T_{\rm H}$ and the parameters $h'(r_h)$, $b/f(r_h)$ as 
functions of $\alpha$ for rotating GB black holes with $\Omega = 0.02$, $Q=1$, $r_h=0.8$
and $\psi\equiv 0$. 
}
\end{figure}

The discussion above suggests that GB black holes exist only up to a maximal value of the GB coupling.
However,  we know that for $\Omega = 0$
black holes also exist for all values of $\alpha$ up to the Chern-Simons limit $\alpha = L^2/4$.
It is therefore a natural question to attempt
to construct the rotating generalizations of these solutions. 
We therefore considered the static solutions close to $\alpha\sim L^2/4$ and constructed rotating generalizations
of these. 
We managed to construct another branch, i.e. a third branch of solutions  in this region of the parameter. 
In contrast to the solutions on the other two branches, the solutions
of this new branch have $h'(r_h) < 0$. In addition, we were able to construct yet another, 
i.e. a fourth, branch of solutions
that coincides with the third one at $\alpha=\tilde{\alpha}$. 

It is worth pointing out  that  the branches of solutions have $h'(r_h) > 0$ and $h'(r_h) < 0$, respectively.
Hence they are completely disjoint. This phenomenon
seems to be specific for charged, rotating solutions in asymptotically AdS. 
Indeed, a similar study in the case of uncharged black hole \cite{Brihaye:2010wx,Brihaye:2011hf} 
reveals the occurrence of
a unique branch of rotating black holes. 

It is also interesting to understand how these solutions behave for large rotation parameter $\Omega$. 
We find that the main branch gets smoothly deformed by the rotation. The third branch
becomes smaller in $\alpha$ and it cannot be extended continuously up to $\alpha = L^2/4$. 
This suggest that $\alpha = L^2/4$ is itself a critical value for rotating black holes and that
charged Chern-Simons black holes cannot rotate. This, however, should be confirmed by an independent analysis
which we do not aim at in this paper. 

We have also studied the influence of the GB term on the solutions.
We find that for small $\alpha$ the black hole terminates into an extremal solution
at a maximal value of $\Omega$. For larger $\alpha$ our numerical results suggest 
that several branches of solutions 
exist that at a critical value of $\Omega$ terminate also into an extremal solution.

\subsubsection{Black holes with scalar hair}
We find that rotating GB black holes with scalar hair exist for smaller values of $r_h$ as compared to the
solutions without scalar hair. 
Our numerical analysis of the solutions for several values of $\Omega$ and $\alpha$ suggests the occurrence 
of a phenomenon similar (although slightly more involved) to the one discussed for black holes without scalar hair.
Choosing $r_h=0.5$ and $Q=1$ we constructed families of
rotating solutions with $\alpha > 0$.

Again, the parameter $h'(r_h)$ plays a crucial role in the understanding of the pattern of solutions. This seems
to diverge for (at least) 
four values of the parameter $\alpha$, say for $\alpha = \tilde{\alpha}_k$, $k=1,2,3,...$. 
By diverging we mean that it tends to $\pm \infty$ for $\alpha$ approaching $\tilde{\alpha}_k$ from the  left and
the right, respectively. 
This critical phenomenon seems to be present already for slowly rotating  black hole and persists 
for larger angular momentum. We observe that
the critical values $\tilde{\alpha}_k$  depend only weakly on the value $\Omega$.
However, when plotting physical quantities as given in Fig. \ref{sardor_05_phys} 
we observe no discontinuities.
We therefore believe that the critical values of $\alpha$ correspond to configurations where the
coordinate gauge fixing $g(r)= r^2 $ becomes accidentally not appropriate to describe the solution. Further study
of this phenomenon is currently under investigation. 

\begin{figure}
\centering
\epsfysize=8cm
\mbox{\epsffile{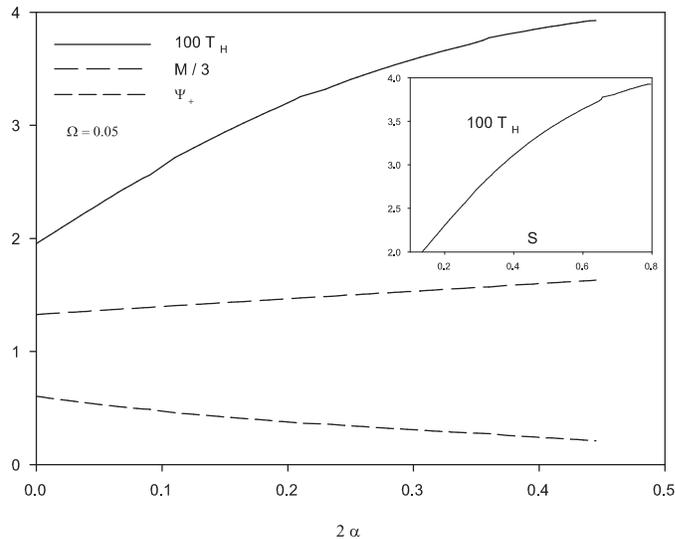}}
\caption{\label{sardor_05_phys}
The mass $M$, temperature $T_H$ and the value of the scalar field on the AdS boundary, $\psi_+$ for the rotating, hairy black holes with $Q=1, r_h=0.5$
and $\Omega = 0.05$ as function of $\alpha$. We also show the temperature $T_{\rm H}$ as function of $S$ in the
subplot.   
}
\end{figure}

 


\clearpage
\section{Conclusions}
In this paper we have studied 5-dimensional solitons and black holes in GB gravity 
coupled to electromagnetic and scalar fields. In the limit of vanishing GB coupling this model
reduces to Einstein-Maxwell theory coupled to a scalar field. 

For vanishing scalar field the static black hole solutions are the RNAdS solution and its GB generalization.
For a fixed value of the electric charge $Q$, these exist for horizon radius larger than
a minimal value. At this minimal value the black hole becomes extremal. Black holes with scalar hair
exist for smaller values of the horizon extending the domain of existence of
the static black hole solutions in the $r_h$-$Q$-plane. When considering the limit $r_h \to 0$
these hairy black hole solutions tend to the corresponding soliton solutions.
\\
In this paper, we have been mainly interested in the rotating generalizations of both types of
solutions. We find that although solitons cannot
be made rotating properly with non-vanishing angular momentum, the hairy black holes 
can be generalized to rotating 
solutions characterized by the angular velocity on the horizon $\Omega$. We find that 
the hairy solutions can rotate up to a maximal value of $\Omega$
where we believe that they become singular.  
\\
Another aspect of our study has been to investigate the effect of the GB term.
In the presence of a negative cosmological 
constant, it is known that static asymptotically AdS solutions exist for $\alpha <  L^2/4$ where $L$
denotes the AdS radius and $\alpha$ the GB coupling constant. The limit $\alpha=L^2/4$ corresponds 
to the Chern-Simons limit.  
Our numerical result show that some branches of soliton solutions disappear
when $\alpha$ is large enough. For the GB black hole solutions we find that these exist for
$0 \leq \alpha \leq \alpha_{cr} < L^2/4$. Moreover, for fixed non-vanishing $\alpha$ several rotating solutions can
exist with different values of $\Omega$ but the same value of the charge $Q$. 

It would be interesting to find analytic arguments for our numerical results. This is currently under investigation.
\\
\\
{\bf Acknowledgments} B.H. and S.T. 
gratefully acknowledge support within the framework of the DFG Research
Training Group 1620 {\it Models of gravity}. Y.B. would like to thank the Belgian F.N.R.S. for financial
support.

\end{document}